\documentclass{aastex63}

\received{XXX}
\revised{YYY}
\accepted{ZZZ}
\usepackage{bm}
\usepackage{amsthm,amsmath,amssymb}
\usepackage{mathrsfs}
\shorttitle{Orbital dynamics}
\shortauthors{Liu et al.}

\begin{document}

\title{Orbital Dynamics with the Gravitational Perturbation due to a Disk}

\correspondingauthor{Xin-Hao Liao}
\email{xhliao@shao.ac.cn, liutao@shao.ac.cn}
\author[0000-0001-9253-1252]{Tao Liu}
\affiliation{Shanghai Astronomical Observatory,Chinese Academy of Sciences,Shanghai 200030,People's Republic of China}
\affiliation{School of Physical Science and Technology,ShanghaiTech University,Shanghai 201210,People's Republic of China }
\affiliation{University of Chinese Academy of Sciences,Beijing 100049,People's Republic of China}

\author{Xue-Qing Xu}
\affiliation{Shanghai Astronomical Observatory,Chinese Academy of Sciences,Shanghai 200030,People's Republic of China}
\affiliation{Key Laboratory of Planetary Sciences,Chinese Academy of Sciences,Shanghai 200030,People's Republic of China}


\author{Xin-Hao Liao}
\affiliation{Shanghai Astronomical Observatory,Chinese Academy of Sciences,Shanghai 200030,People's Republic of China}
\affiliation{Key Laboratory of Planetary Sciences,Chinese Academy of Sciences,Shanghai 200030,People's Republic of China}


\begin{abstract}

The secular behavior of an orbit under the gravitational perturbation due to a two-dimensional uniform disk is studied in this paper, through analytical and numerical approaches. We develop the secular approximation of this problem and obtain the averaged Hamiltonian for this system first. We find that, when the ratio of the semimajor axes of the inner orbit and the disk radius takes very small values ($\ll1$), and if the inclination between the inner orbit and the disk is greater than the critical value of $30^\circ$, the inner orbit will undergo the (classical) Lidov-Kozai resonance in which variations of eccentricity and inclination are usually very large and the system has two equilibrium points at $\omega=\pi/2,3\pi/2$ ($\omega$ is the argument of perihelion). The critical value will slightly drop to about $27^\circ$ as the ratio increases to 0.4. However, the secular resonances will not occur for the outer orbit and the variations of the eccentricity and inclination are small. When the ratio of the orbit and the disk radius is nearly $1$, there are many more complicated Lidov-Kozai resonance types which lead to the orbital behaviors that are different from the classical Lidov-Kozai case. In these resonances, the system has more equilibrium points which could appear at $\omega=0,\pi/2,\pi,3\pi/2$, and even other values of $\omega$. The variations of eccentricity and inclination become relatively moderate, moreover, in some cases the orbit can be maintained at a highly inclined state. In addition, a analysis shows that a Kuzmin disk can also lead to the (classical) Lidov-Kozai resonance and the critical inclination is also $30^\circ$.

\end{abstract}

\keywords{celestial mechanics --- orbital evolution --- orbital resonances ---  protoplanetary disks --- galaxy disks }


\section{Introduction} \label{sec:intro}

Disk-like structures are ubiquitous in the universe, from the galactic disks in spiral galaxies to the accretion disks in active galactic nuclei (AGN) and X-ray binaries, to protoplanetary disks, debris disks, and planetary rings in the formation and evolution process of planetary systems \citep{sellwood_1989,latter2018planetary}. In our solar system, all outer planets have planetary rings and the disk-shaped Kuiper belt has been observed. In addition, there is a theoretical Oort cloud considered as comet reservoir in the outer solar system, including a disk-shaped inner Oort cloud and a spherical outer Oort cloud \citep{oort1950structure,hills1981comet,levison2001origin}, which has not yet been confirmed by observation. Recently, \citet{sefilian2019shepherding} proposed that a debris disc of icy material exists outside the orbit of Neptune, with a combined mass around 10 times that of Earth, whose self-gravity could be responsible for the strange orbital architecture of Trans-Neptunian Objects (TNOs) in the outer solar system \citep{trujillo2014a,batygin2016evidence}.

The gravitational effect of a astronomical disk plays an important role in the formation of the dynamical architecture of various systems. \citet{nagasawa2003eccentricity} showed that the self-gravity of a dissipative protoplanetary disk could have a significant impact on the planetary eccentricity in the extrasolar multiple planetary system. In binary systems, many exoplanets have higher eccentricities than that of the planets in the solar system, some of which could be induced via the Lidov-Kozai effect (or resonance; mechanism) \citep{lidov1962evolution,kozai1962secular,holman1997chaotic,innanen1997kozai,wu2003planet,takeda2005high}. However, when the inclination angle between the protoplanetary disk plane, in which planetesimals are embedded, and the binary plane is too large, the growth of the kilometer-sized planetesimals could be inhibited due to the Lidov-Kozai effect \citep{marzari2009planet}. This is a challenge for the current planetary formation theories. In order to understand planet formation in stellar binaries, many authors have investigated the influence of the protoplanetary disk gravity on planetesimal dynamics. Some found that the fast apsidal precession on planetesimal orbit induced by the gravitational effect of an axisymmetric protoplanetary disk can effectively suppress the Lidov-Kozai effect or the excitation of planetesimal eccentricity, which is conducive to the planetesimal growth resulting in the formation of planetary embryo \citep{batygin2011formation,rafikov2013building,rafikov2013planet}. However, it was shown that the gravity of an eccentric disk will instead excite planetesimal eccentricity to high values, leading to high impact velocities, and therefore prevent their growth \citep{marzari2012eccentricity,marzari2013influence,silsbee2014planet,rafikov2014planet,lines2016modelling}. \citet{zhao2012planetesimal} studied the Lidov-Kozai effect on planetesimal dynamics with the perturbations from both the companion star and the circumprimary disk in the inclined binary system. They showed that the Lidov-Kozai effect will be similarly suppressed if the gravitational effect of disk is included, but the Lidov-Kozai effect can work at arbitrarily low inclinations in the Kozai-on region in which planetesimal eccentricities can be excited to extremely high values ($\sim1$). Hence the planetesimal with very high orbital eccentricity may become a ``hot planetesimal" as the shrink of the planetesimal orbit due to the gas drag damping of the gaseous disk.

\citet{terquem2010eccentricity} found that the secular perturbation of an annulus disk will lead to the Lidov-Kozai effect for a planet if the planetary orbit is well inside the disk inner cavity. Furthermore, if the planetary orbit crosses the disk but most of the disk mass is beyond the orbit, they found that the oscillations of both the planetary eccentricity and inclination were not observed when the initial inclination of the orbit is below a critical value, which is significantly smaller than 39.2$\degr$. The authors pointed out that the critical value could be 30$\degr$ in some case (case A in the paper). The analogous discussion for a three-dimensional disk and a warped disk can be found in the literature  \citet{teyssandier2012orbital} and \citet{terquem2013effects}, respectively.

In galactic dynamics, a series of papers \citep{vokrouhlicky1998stellar,vsubr2004star,vsubr2005highly,karas2007enhanced} investigated the secular evolution of the stellar orbits in a galactic center surrounded by a massive accretion disk. They indicated that the stellar orbits near the black hole will undergo the Lidov-Kozai resonance under the perturbation of the accretion disk. This is helpful for the formation of highly eccentric stellar orbit in the vicinity of the black hole and increasing the star-capture rate of the black hole. Hass and \v{S}ubr also studied the Lidov-Kozai resonance in an eccentric stellar disk around a supermassive black hole \citep{haas2016rich,vsubr2016properties}.

In this paper, we aim to understand the orbital dynamics under the secular perturbation of a disk through both analytical and numerical methods. We analytically demonstrate for the first time that the gravitational perturbation from a uniform disk will induce the Lidov-Kozai effect (or resonance). In Section \ref{sec:model}, we describe the dynamical model of a massless test particle under the disk perturbation, focusing on the multipole expansion of the disturbing function and its averaging. In Section \ref{sec:qualitative}, we provide a analytic study on the secular problem. Successive numerical studies are presented in Section \ref{sec:numerical}. And results are summarize and discussed in Section \ref{sec:discussion}.
\section{THE DYNAMICAL MODEL} \label{sec:model}
We consider a massless test particle orbiting around a central body of mass $M_{\star}$ which is surrounded by a disk of mass $m_d$. We assume $m_d\ll M_{\star}$, hence the motion of the particle is dominated by the central body, and the particle's orbit is a Keplerian ellipse but slightly perturbed by the gravitational potential of the disk. The Hamiltonian for this system is written as follows:
\begin{equation}
F=\frac{\mu}{2a}-V
\label{con:eq1}
\end{equation}
where $\mu=\mathcal{G}M_{\star}$ ($\mathcal{G}$ is the gravitational constant), $a$ is the semimajor axis of the particle's orbit; $V$ is the gravitational potential of the disk. Note that the Hamiltonian has the opposite sign relative to the standard form. For simplicity, in our case the disk is considered to be a two-dimensional uniform disk. The gravitational potential exerted by the uniform disk on the particle is given by \citep{alberti2007dynamics}
\begin{equation}
V=-2\mathcal{G}\sigma\int_{0}^{R}\int_{0}^{\pi}\frac{\rho d\rho d\theta}{\sqrt{r^2+\rho^{2}-2\rho r\cos \theta \cos \varphi}}
\label{con:eq2}
\end{equation}
where $R$ and $\sigma$ are the radius and constant mass density of the disk, respectively. $r$ is the distance between the particle and the disk's center (or the central body), $\varphi$ is the angle between the position vector $\boldsymbol{r}$ of the particle and the disk plane.

Since we are interested in the secular behaviour of the particle's orbit under the perturbation from the disk, we would like to average the perturbing potential $V$ (or the Hamiltonian) over the mean anomaly $M$ of the particle's orbit, and this results in the elimination of the short-period terms in the perturbing potential. This process is know as secular approximation. Unfortunately, it is almost impossible to average Equation (\ref{con:eq2}) directly and then obtain an analytical expression even for a constant mass density $\sigma$. However, when $r/R<1$ (or $r/R>1$), Equation (\ref{con:eq2}) can be expanded in $r/R$ (or $R/r$) by means of Legendre's polynomials $P_n$. This result in

\noindent$r/R<1:$
\begin{equation}
V=-\frac{2\mathcal{G}m_d}{R}\left\{1-\frac{r}{R}\sin\varphi+\frac{1}{4}\left(\frac{r}{R}\right)^{2}(3\sin^2\varphi-1)+O\bigg(\left(\frac{r}{R}\right)^3\bigg)\right\}
\label{con:eq3}
\end{equation}
where $m_d=\sigma\pi R^2$, and

\noindent$r/R>1:$
\begin{equation}
V=-\frac{\mathcal{G}m_d}{R}\left\{\frac{R}{r}-\frac{1}{4}\left(\frac{R}{r}\right)^{3}\left(1-\frac{3}{2}\cos^2\varphi\right)+O\bigg(\left(\frac{R}{r}\right)^5\bigg)\right\}
\label{con:eq4}
\end{equation}
The detailed derivation of Equations (\ref{con:eq3}),(\ref{con:eq4}) can be seen in Appendix \ref{sec:A1}.

We take the disk plane to be the equatorial plane of the central body, and the inclination of the particle's orbit is measured with respect to this plane. Thus, we have
\begin{equation}
\sin\varphi=\sin i\cdot|\sin(f+\omega)|
\end{equation}
where $i,\omega,f$ are the inclination, argument of perihelion, and true anomaly of the particle's orbit respectively (moreover, we use the most common variables $e,\Omega$ to denote the eccentricity and longitude of ascending node of the orbit in this paper). In fact, the key step in the process of averaging Equation (\ref{con:eq3}) is to obtain the average value of the formula $r |\sin(f+\omega)|$, the average value and details are presented in Appendix \ref{sec:A2}.

Finally, averaging the potential in Equations (\ref{con:eq3}),(\ref{con:eq4}) over the mean anomaly $M$, we get

\noindent$r/R<1:$
\begin{equation}
\begin{aligned}
\overline{V}=\frac{\mathcal{G}m_d}{R}&\left\{\left(\frac{a}{R}\right)\frac{4\sin i}{\pi}\left[\left(1+\frac{1}{2}e^2\right)-e^2\cos2\omega\right] \right.\\
&+\left.\left(\frac{a}{R}\right)^2\left[\frac{1}{2}\left(1+\frac{3}{2}e^2\right)\left(1-\frac{3}{2}\sin^2i\right)+\frac{15}{8}e^2\sin^2i\cos2\omega\right]\right\}
\end{aligned}
\label{con:eq8}
\end{equation}
We drop the constant term independent of the orbital elements in the above expansion. The above expansion is at the quadrupole level of approximation. The first (second) term in the expansion is called the dipole (quadrupole), which is proportional to $a/R$ ($(a/R)^2$). Note that Equation (\ref{con:eq8}) only applies to the inner orbits whose apocenter distances are smaller than the disk radius $R$ (geometrically, the inner orbit is located inside the sphere of radius $R$). Likewise

\noindent$r/R>1:$
\begin{equation}
\overline{V}=-\frac{\mathcal{G}m_d}{R}\left\{\frac{R}{a}+\frac{1}{8}\left(\frac{R}{a}\right)^3\left(1-\frac{3}{2}\sin^2i\right)(1-e^2)^{-3/2}\right\}
\label{con:eq9}
\end{equation}
Equation (\ref{con:eq9}) only applies to the outer orbits whose pericenter distances are greater than the disk radius $R$. The outer orbit is outside the sphere of radius $R$.

A closed form of the potential of uniform disk had been derived in \citet{lass1983gravitational}, involving complete elliptic integrals of three kinds \citep{byrd1971handbook}. The closed form is numerically equivalent to the integral form in Equation (\ref{con:eq2}), and Equations (\ref{con:eq3})(\ref{con:eq4}) can also been obtained by expanding the closed form in the appropriate limits, but the derivations are very complicated and tedious (particularly for the case of $r/R>1$). On the other hand, complete elliptic integrals as well as the potential in closed form can be computed precisely and fast, in comparison with the integral form, hence we will adopt the closed form in our full model where the potential of uniform disk is neither approximated nor averaged (see details in Section \ref{sec:numerical}).
\section{QUALITATIVE ANALYSIS OF SECULAR DYNAMICAL BEHAVIOR} \label{sec:qualitative}

In this section we star the qualitative study of the dynamics of a particle's orbit under the secular perturbation from the uniform disk. We consider two secular problems: The first is about the inner orbit and the second is about the outer orbit.
\subsection{Dynamics for the inner orbit} \label{subsec:case1}
It is convenient to understand the dynamical behavior of the orbit using the canonical Delaunay variables \citep{brouwer1961methods}:
\begin{equation}
\left\{\begin{aligned}
&L=\sqrt{\mu a}           &l=M\\
&G=\sqrt{\mu a(1-e^2)}    &g=\omega\\
&H=\sqrt{\mu a(1-e^2)}\cos i    &h=\Omega
\end{aligned}
\right.
\label{con:eq10}
\end{equation}
In the inner orbit problem, the averaged Hamiltonian for the system considered here is
\begin{equation}
\overline{F}=\frac{\mu ^2}{2L^2}-\overline{V}
\label{con:eq11}
\end{equation}
with
\begin{equation}
\begin{aligned}
&\overline{V}=\overline{V}_{di}+\epsilon \overline{V}_{quad}\\
&\overline{V}_{di}=k\left(\frac{a}{R}\right)\left[\left(1+\frac{1}{2}e^2\right)-e^2\cos2\omega\right]\sin i\\
&\overline{V}_{quad}=k\left(\frac{a}{R}\right)^2\frac{\pi}{8}\left[\left(1+\frac{3}{2}e^2\right)\left(1-\frac{3}{2}\sin^2i\right)+\frac{15}{4}e^2\sin^2i\cos2\omega\right]
\end{aligned}
\label{con:eq11}
\end{equation}
where $k=(4\mathcal{G}m_d /\pi R)$. $\overline{V}_{di}$ and $\overline{V}_{quad}$ are the dipole term and the quadrupole term of the potential $\overline{V}$ in Equation (\ref{con:eq8}), respectively. $\epsilon$ is a dimensionless control parameter which takes a value of $0$ $(1)$ in the dipole (quadrupole) approximation of the potential/Hamiltonian.

The averaged Hamiltonian does not depend on the mean anomaly $M$, nor the longitude of ascending node $\Omega$, thus
\begin{equation}
\dot{L}=\frac{\partial \overline{F}}{\partial l}=0,\:\:\:\dot{H}=\frac{\partial \overline{F}}{\partial h}=0
\label{con:eq13}
\end{equation}
$L$,$H$ are constant of motion, which implies that the semimajor axis $a$, the $z$ component of the angular momentum of the orbit are conserved in the secular problem. Apparently, $H/L$ also remains constant, it follows that
\begin{equation}
\sqrt{1-e^2}\cos i=J_z
\label{con:eq14}
\end{equation}
where $J_z$ is a constant (Kozai integral). This indicates oscillations of $e$ and $i$ are coupled and in antiphase. Since $L$,$H$ and the Hamiltonian itself are all constant, the degree of freedom for this system is reduced to one, related to the couple ($G,g$). Thus the system is analytically integrable in principle.

The equations of motion about the canonical variables $G,g$ are given by
\begin{equation}
\dot{G}=\frac{\partial \overline{F}}{\partial g}=-k\left(\frac{a}{R}\right)\left(2-\epsilon\frac{15\pi}{16}\frac{a}{R}\sin i\right)e^2\sin i\cdot\sin2\omega
\label{con:eq15}
\end{equation}

\begin{equation}
\begin{aligned}
\dot{g}=-\frac{\partial \overline{F}}{\partial G}=-\frac{k}{G}&\left\{\left(\frac{a}{R}\right)\left[(1-e^2)(1-2\cos2\omega)\sin i-\frac{\cos^2i}{\sin i}\left(1+\frac{1}{2}e^2-e^2\cos2\omega\right)\right]\right.\\
&-\left.\epsilon\left(\frac{a}{R}\right)^2\frac{3\pi}{16}\bigg[1-e^2-5\cos^2i+5\left(e^2-\sin^2i\right)\cos2\omega\bigg]\right\}
\label{con:eq013}
\end{aligned}
\end{equation}
When $a$ is much smaller than $R$ (i.e., $a/R\ll1$), the quadrupole term effect is negligible and hence the dipole approximation can describe the dynamical behaviour of the system well.
Next, we first consider the above equations of motion in the dipole approximation.
\subsubsection{Dipole approximation $(\epsilon=0)$} \label{subsubsec:case11}
The equilibrium point of the system satisfies the following equations
\begin{equation}
\left\{\begin{aligned}
&\dot{G}=\frac{\partial \overline{F}}{\partial g}=0\\
&\dot{g}=-\frac{\partial \overline{F}}{\partial G}=0
\end{aligned}
\label{con:eq17}
\right.
\end{equation}
In the dipole approximation ($\epsilon=0$), solving $\dot{G}=0$, then we get: $\omega=0 ,\pi/2,\pi,3\pi/2$. Substitution of these values of $\omega$ into Equation (\ref{con:eq013}) yields
\begin{equation}
\dot{g}>0,\:\:\:\text{at}\:\:\omega=0,\pi\:\:(\cos2\omega=1)
\end{equation}
Thus Equations (\ref{con:eq17}) has no solution at $\omega=0,\:\pi$. At $\omega=\pi/2,\:3\pi/2\:\:  (\cos2\omega=-1)$, Equation (\ref{con:eq013}) becomes
\begin{equation}
\dot{g}=-\frac{k}{G\sin i}\left(\frac{a}{R}\right)\left\{3\frac{G^2}{L^2}-\frac{3}{2}\frac{H^2}{L^2}-\frac{5}{2}\frac{H^2}{G^2}\right\}
\end{equation}
The above equation is expressed in terms of the Delaunay variables. As mentioned above, where $L$ as well as $H$ are constant. Solving the equation $\dot{g}=0$, one obtains
\begin{equation}
G^2=\frac{3H^2+\sqrt{9H^4+120L^2H^2}}{12}
\end{equation}
As $G^2\leq L^2$, it follows that
\begin{equation}
\left|\frac{H}{L}\right|\leq\frac{\sqrt{3}}{2}
\end{equation}
namely
\begin{equation}
\left|\sqrt{1-e^2}\cos i\right|\leq\frac{\sqrt{3}}{2} \:\:\: \text{or} \:\:\: |J_z|\leq\frac{\sqrt{3}}{2}
\label{con:eq25}
\end{equation}

Therefore, when the above inequality is satisfied, $\dot{g}=0$ as well as Equations (\ref{con:eq17}) have solutions, and the system has two equilibrium points at $\omega=\pi/2,3\pi/2$. This implies that the Lidov-Kozai resonance (or effect) will occur for the inner orbits of $|J_z|\leq\sqrt{3}/2$. When $|J_z|>\sqrt{3}/2$, Equations (\ref{con:eq17}) do not have solution and hence there is no equilibrium point for the system. The inner orbits of $|J_z|>\sqrt{3}/2$  do not undergo any secular resonances. Apparently, the critical value of $J_c$ for the occurrence of the Lidov-Kozai resonance is $\sqrt{3}/2$ (for the prograde orbits), i.e. $J_c=\sqrt{3}/2$.

\begin{figure*}[h]
\centering
\gridline{\fig{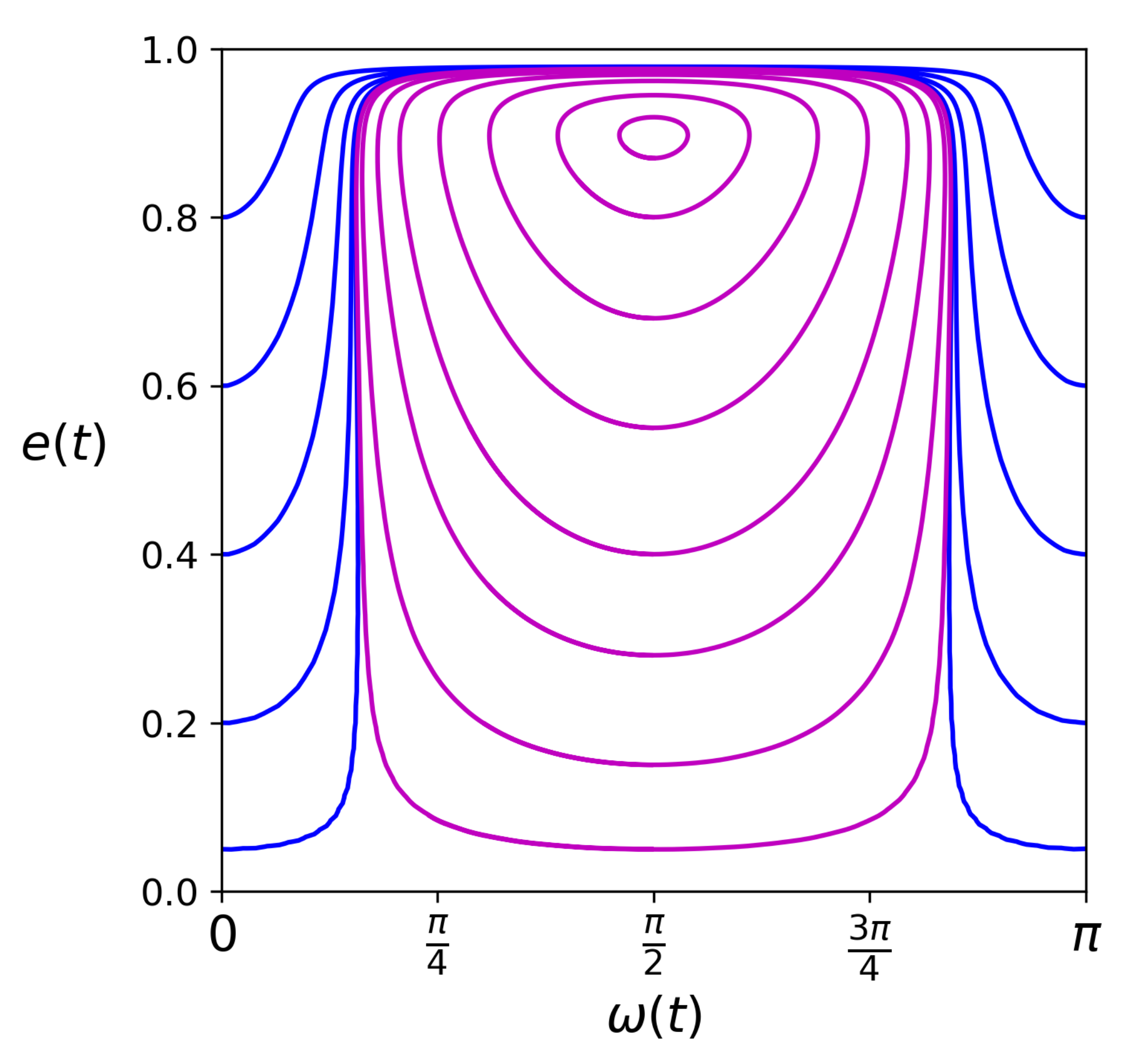}{0.27\textwidth}{(a)}
           \hspace{-5cm}
          \fig{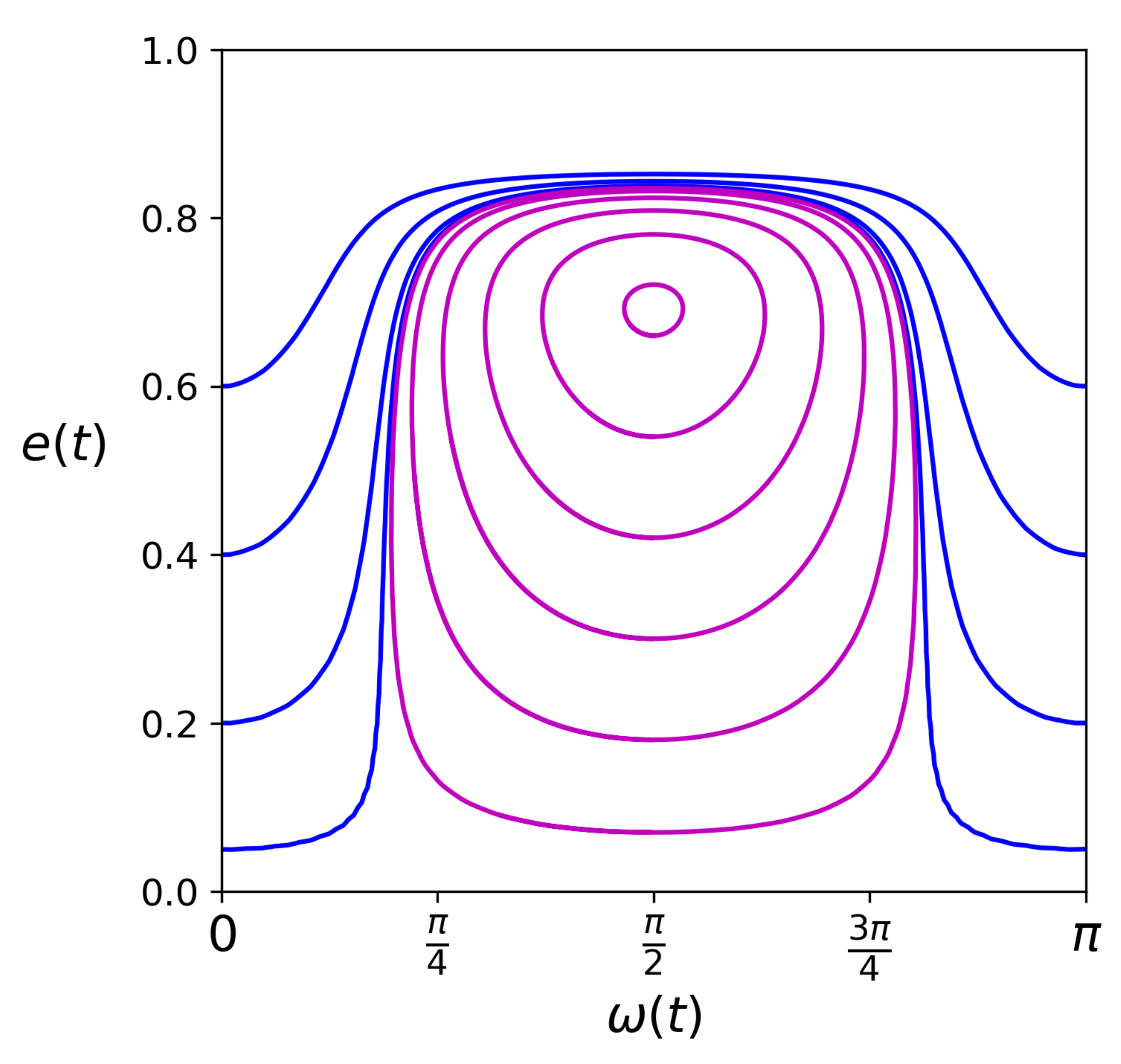}{0.27\textwidth}{(b)}
          }
\gridline{\fig{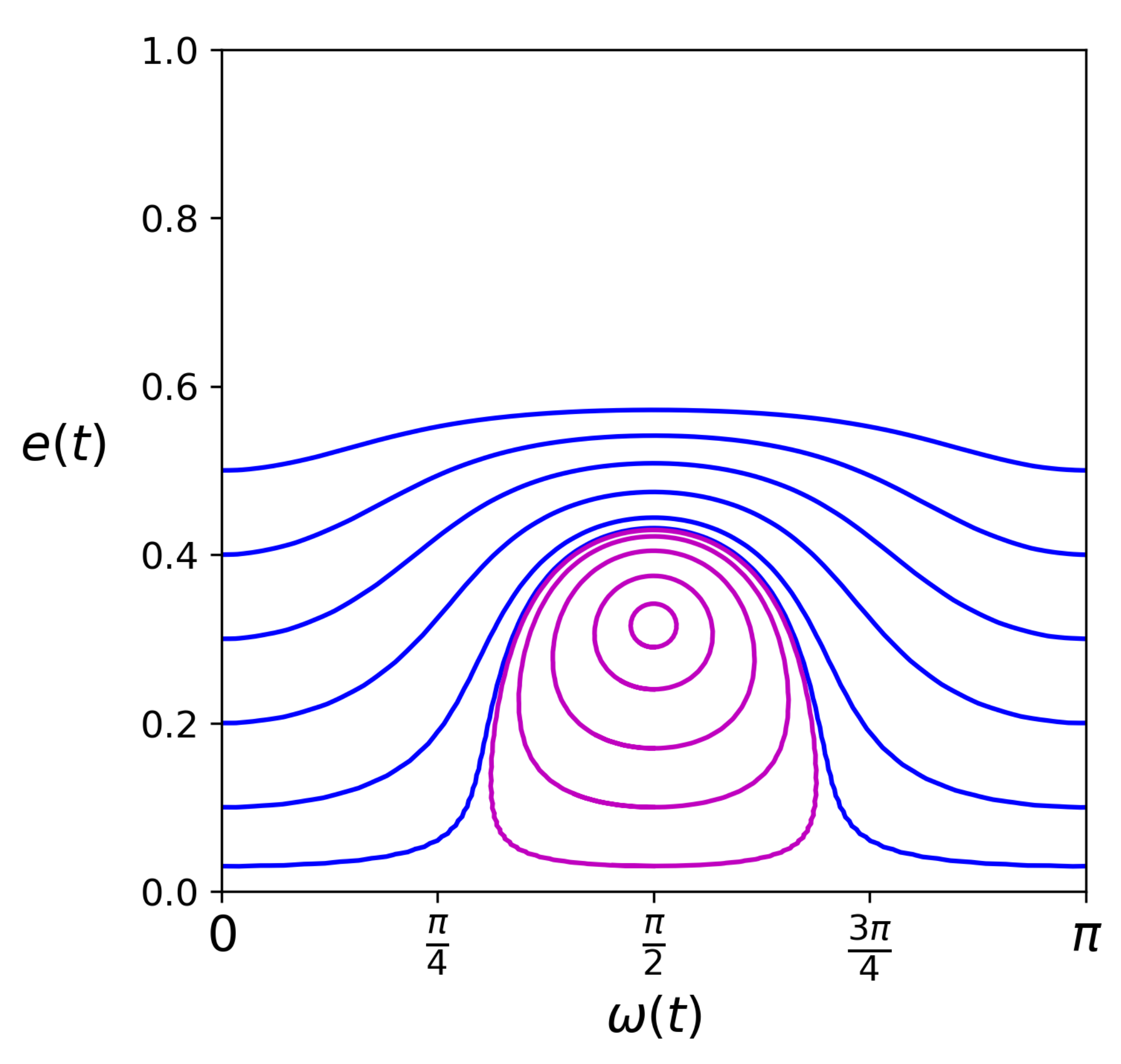}{0.27\textwidth}{(c)}
          \hspace{-5cm}
          \fig{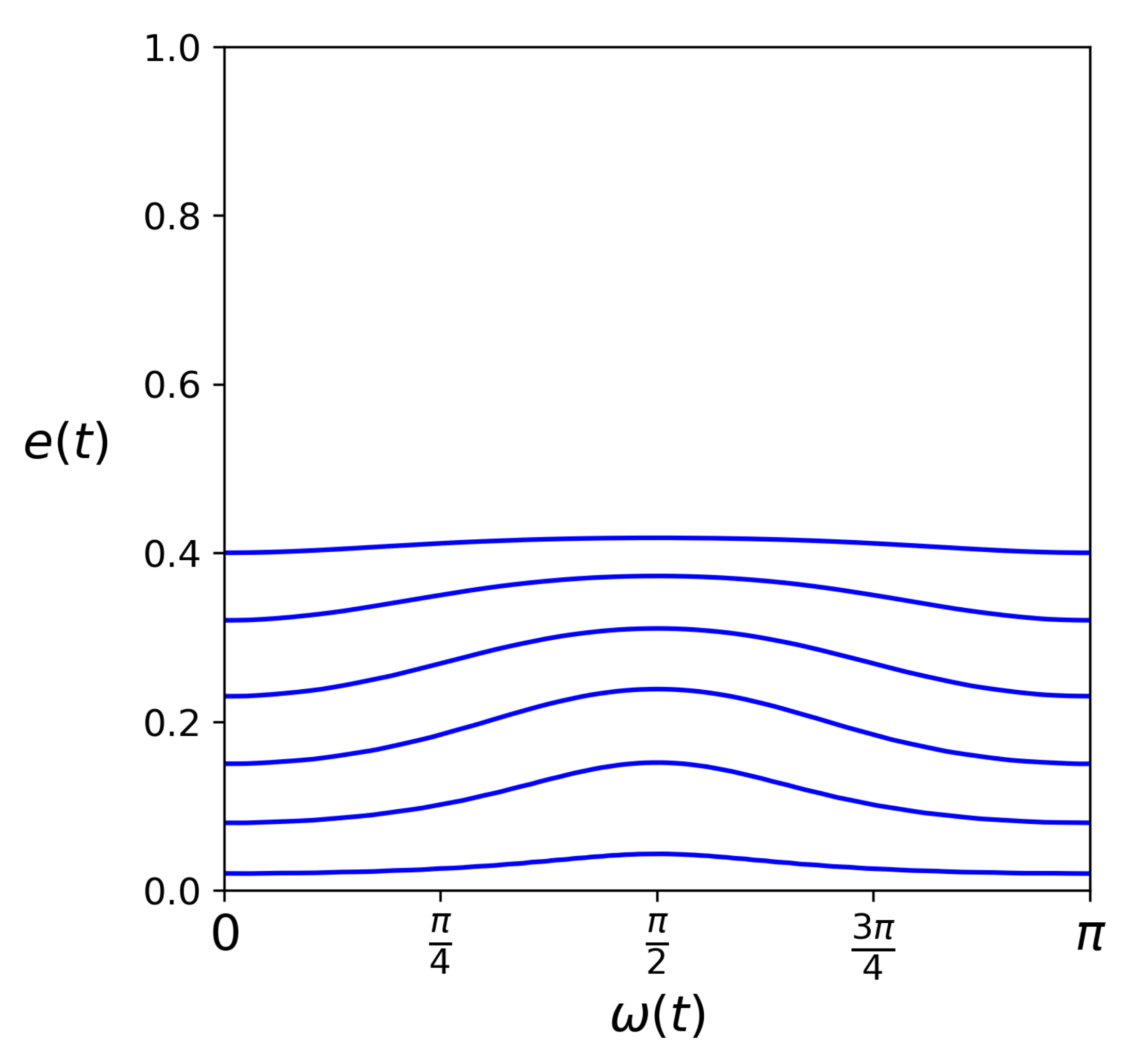}{0.27\textwidth}{(d)}
          }
\caption{$(e,\omega)$ phase space trajectories in the dipole approximation. In the calculations, we set: $\mathcal{G}=1,M_{\star}=1, m_d=0.01,R=100$. And we take $a=10$ for all orbits. Where: (a)$J_z=0.2$; (b) $J_z=0.5$; (c) $J_z=0.8$; (d) $J_z=0.9$. The phase space trajectories in $\omega \in(\pi,2\pi)$ are identical to those in $(0,\pi)$, and they are not be presented here. The blue and magenta lines represent the circulating trajectories in $(0,2\pi)$ and librating trajectories around $\omega=\pi/2$, respectively.}
\label{fig:fig2}
\end{figure*}

Figure \ref{fig:fig2} shows the trajectories in the ($e,\omega$) phase space for the inner orbits with different values of $J_z$. In our numerical calculations, we take such a set of dimensionless parameters: $\mathcal{G}=1,M_{\star}=1, m_d=0.01,R=100$ (hereinafter the same). In Figure \ref{fig:fig2}, all orbits have the semimajor axis $a=10$. Figure \ref{fig:fig2}(a)-(c) correspond to $J_z=0.2,0.5,0.8$ respectively. In Figure\ref{fig:fig2}(a)-(c), the Lidov-Kozai resonance occurs, and there is a stable equilibrium point at $\omega=\pi/2$ surrounded by the libration island (closed trajectories). Variations of $e$ are usually very large when the Lidov-Kozai resonance is active, and small eccentricity can even be excited to near 1 for $J_z=0.2$ (see Figure \ref{fig:fig2}(a)). Moreover, it is predictable that variations of $i$ are also usually very large in the Lidov-Kozai resonance, because $\sqrt{1-e^2}\cos i$ remains constant and $e,i$ oscillate in anti-phase. In Figure \ref{fig:fig2}(d), $J_z=0.9\:(>\sqrt{3}/2)$, the Lidov-Kozai resonance does not occur for the orbits, there is no equilibrium point and variations of $e$ are small.

\begin{figure}[ht]
\centering
\includegraphics[scale=0.2]{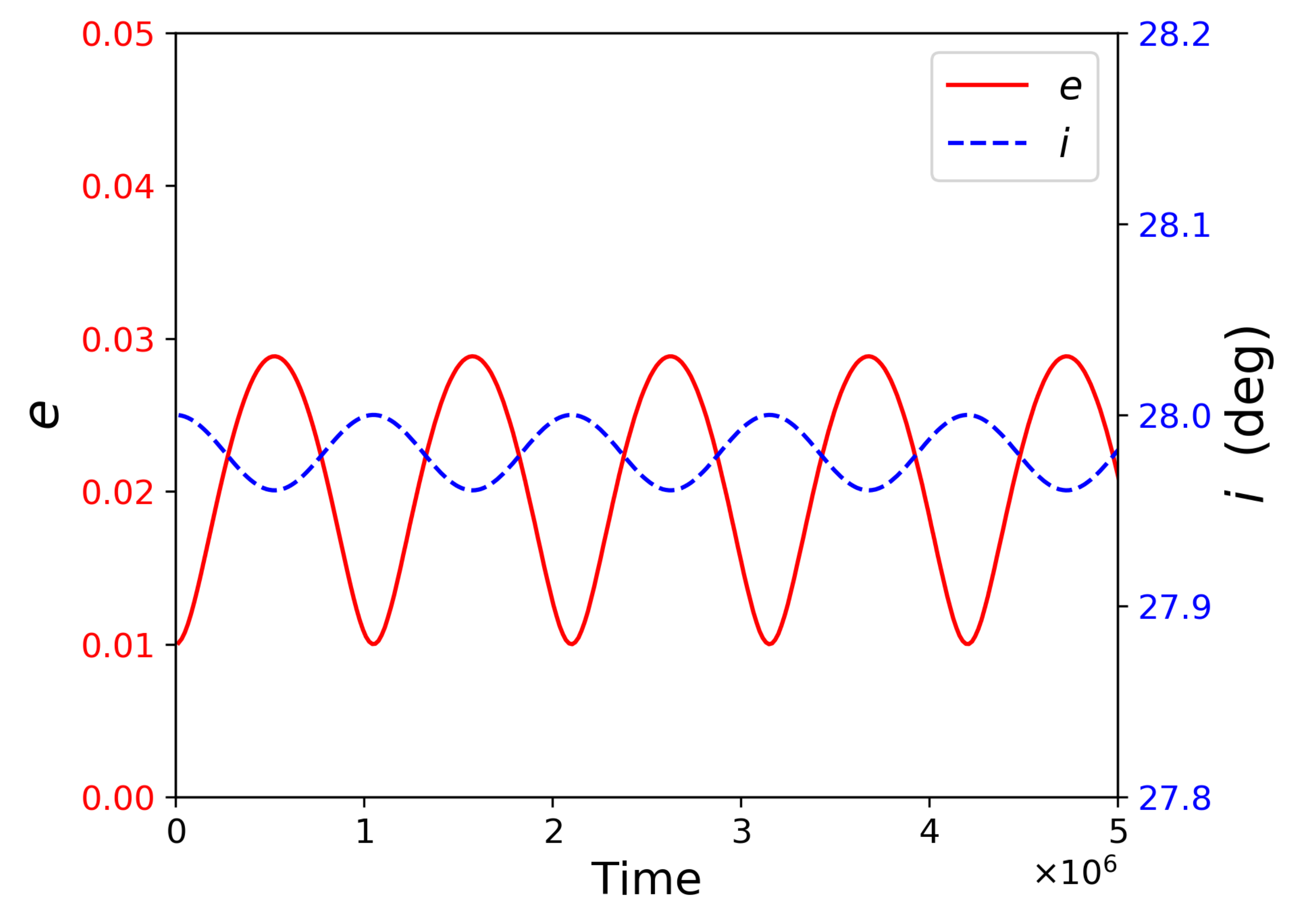}
\includegraphics[scale=0.2]{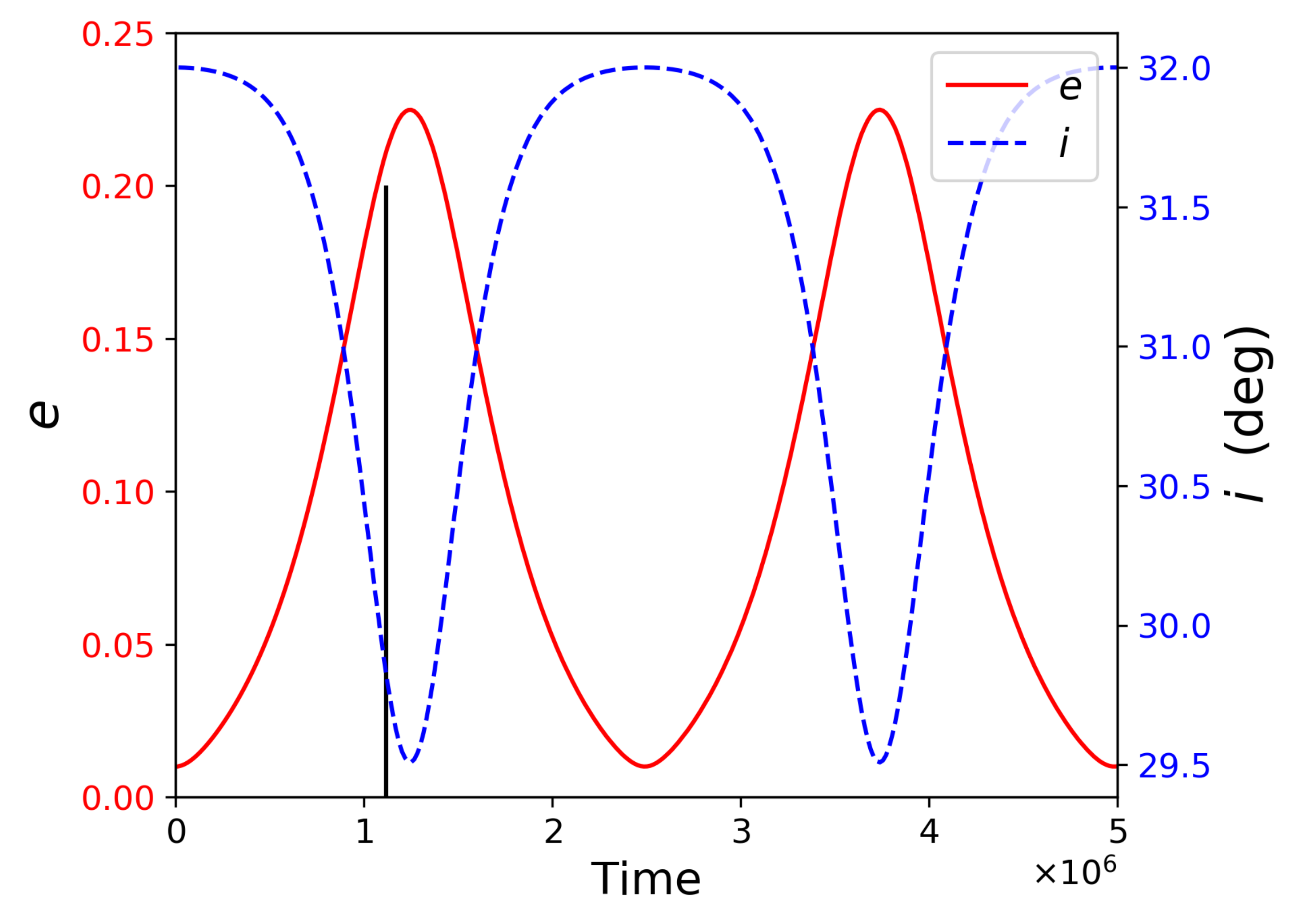}
\includegraphics[scale=0.2]{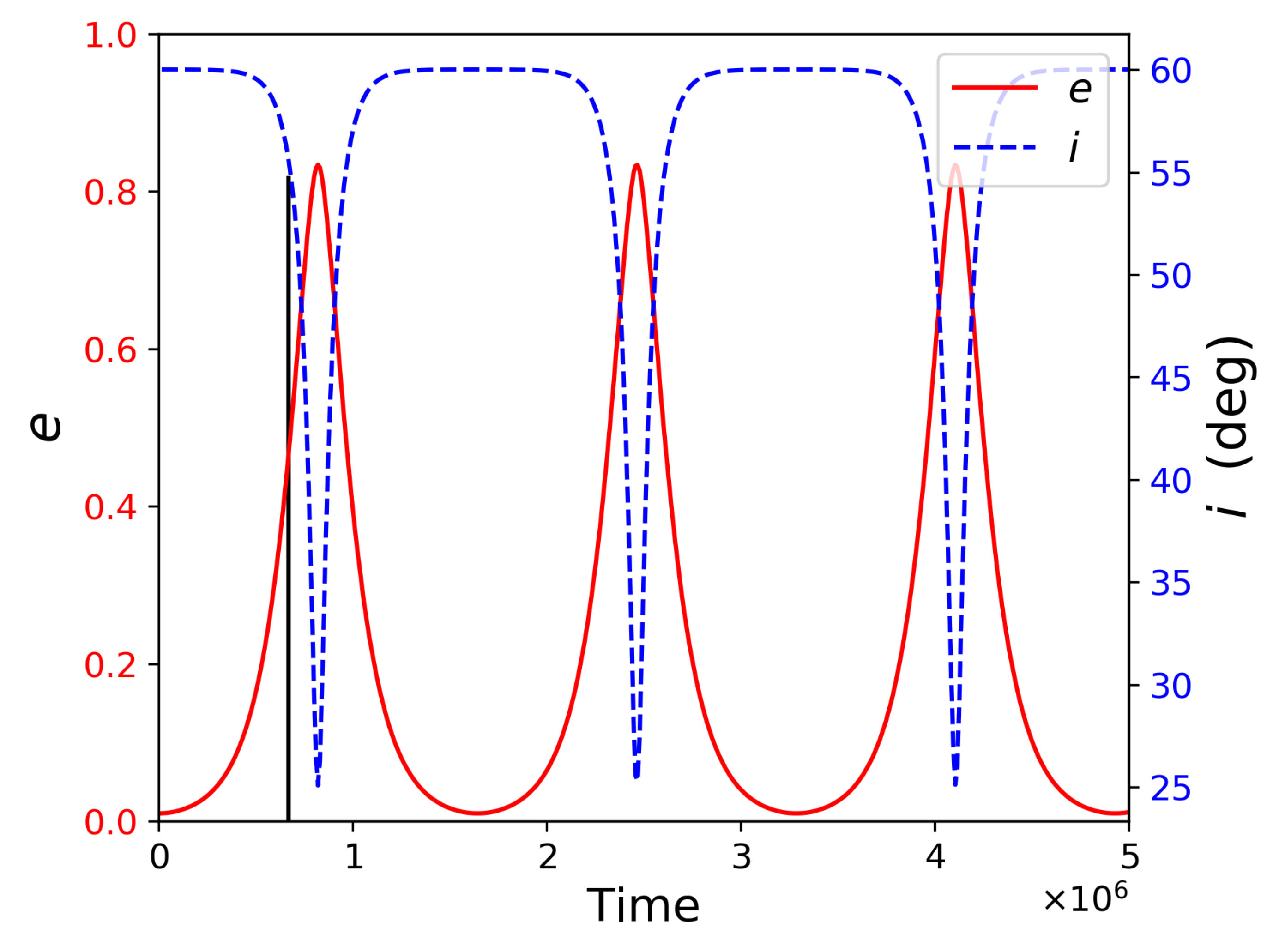}
\caption{Oscillations of $e$,$i$ over time for the orbits with different initial inclinations in the dipole approximation. The orbits in all panels start with the same initial orbital elements: $a_0=10$, $e_0=0.01$, but the left panel with the initial inclination $i_0=28^\circ$, the middle panel: $i_0=32^\circ$, and the right panel: $i_0=60^\circ$. The red solid line and blue dashed line represent the curves of $e$ and $i$ respectively. The vertical lines illustrate the theoretical values of $T_{evol}$ and $e_{max}$ given by Equations (\ref{con:eq0029}),(\ref{con:eq030}).}
\label{fig:fig3}
\end{figure}

The condition making Inequality (\ref{con:eq25}) to hold for any $e$ is:
\begin{equation}
|\cos i|\leq\frac{\sqrt{3}}{2}\:\:\:\text{or}\:\:\:30\degr\leq i\leq 150\degr
\label{con:eq26}
\end{equation}
In other words, as long as the inclination of the inner orbit is larger than 30$\degr$, the orbit will undergo the Lidov-Kozai effect in which both $e$ and $i$ oscillate dramatically. When the inclination is below 30$\degr$, the Lidov-Kozai effect does not work and the oscillations of both $e$ and $i$ are very small. Figure \ref{fig:fig3} shows the coupled oscillations of $e$ and $i$ for the orbits with the initial inclination angels $i_0=28^\circ,32^\circ,60^\circ$. For $i_0=28^\circ$ (left panel), $e$ oscillates between 0.01 and 0.03, and the oscillation of $i$ does not exceed $0.1^\circ$. The amplitudes of $e,i$ are small. However, when $i_0$ is slightly increased to $32^\circ$, $e$ is excited to 0.22 from 0.01 due to the Lidov-Kozai effect (middle panel). For $i_0=60^\circ$ (right panel), the Lidov-Kozai effect becomes very significant, $e$ dramatically oscillates between 0.01 and 0.83, whereas $i$ between $60^\circ$ and $25^\circ$ (approximately). The oscillations of both $e$ and $i$ are very large.

The Lidov-Kozai effect in the disk problem is similar to that in the restricted
three-body problem in nature. The difference is that the critical inclination in the disk problem is 30$\degr$ (in the dipole approximation), and in the restricted
three-body problem is 39.2$\degr$ \citep{kozai1962secular,innanen1997kozai,naoz2013secular}.

Following the analysis of \cite{innanen1997kozai} for the Lidov-Kozai effect in the restricted three-body problem, one can obtain the maximum value reached by the eccentricity and the evolution time $T_{evol}$ to reach the maximum value starting from a small initial eccentricity $e_0$ for the disk problem. We briefly present here the derivation. According to Equations (\ref{con:eq14}),(\ref{con:eq15}), we have
\begin{equation}
\begin{aligned}
&\frac{de}{dt}=2\alpha e\sqrt{1-e^2}\sin i\cdot \sin2\omega \\
&\frac{di}{dt}=2\alpha \frac{e^2}{\sqrt{1-e^2}}\cos i\cdot\sin2\omega
\end{aligned}
\label{con:eq025}
\end{equation}
where $\alpha=(k\sqrt{a/\mu})/R$. For the orbit with very small initial eccentricity $e_0$ and large initial inclination $i_0$, the $i$ remains almost constant before $e$ is excited to a large value because $di/dt$ has the factor $e^2$. And when the Lidov-Kozai effect works, $\omega$ will quickly move to a value which makes $\dot{\omega}=0$, thus according to Equation (\ref{con:eq013}) we have $2\sin^2i_0(1-\cos2\omega)=1$. Taking only the first order of $e$, $de/dt$ becomes
\begin{equation}
\frac{de}{dt}=2\alpha\:e\sin i_0\cdot \sin2\omega=\alpha e\frac{\sqrt{4\sin^2i_0-1}}{\sin i_0}
\label{con:eq028}
\end{equation}
Solving Equation (\ref{con:eq028}), one gets the time $T_{evol}$ it takes to reach $e_{max}$ starting from $e_0$ by

\begin{equation}
T_{evol}=\tau\ln{\left(\frac{e_{max}}{e_0}\right)}\frac{\sin i_0}{\sqrt{4\sin^2i_0-1}}
\label{con:eq0029}
\end{equation}
with the time scale $\tau$
\begin{equation}
\tau=\frac{R^2}{8a^2}\frac{M_{\star}}{m_d}T
\end{equation}
where $T$ is the orbital period. Note that we must have $4\sin^2i_0>1$, namely $i_0>30^\circ$, for the increase of $e$, which is consistent with the previous analysis (Equation (\ref{con:eq26})). If the initial inclination is smaller than $30^\circ$, the actual growth of eccentricity is very small.

For very small initial eccentricity $e_0$ and large initial inclination $i_0$($>30^\circ$), since $\sqrt{1-e^2}\cos i$ remains constant, the eccentricity grows from $e_0$ to $e_{max}$ simultaneously as the inclination drops from $i_0$ to $i_{min}$, that is
\begin{equation}
\sqrt{1-e_{max}^2}\cos i_{min}=\sqrt{1-e_{0}^2}\cos i_{0}
\end{equation}
According to Equation (\ref{con:eq028}), the minimum value $i_{min}$ the inclination can drop to is $30^\circ$. Thus, ignoring the small quantity of $e_0^2$, one obtains
\begin{equation}
e_{max}=\sqrt{1-\frac{4}{3}\cos^2i_0}
\label{con:eq030}
\end{equation}

Two examples illustrating the values of $T_{evol}$ and $e_{max}$ predicted by Equations (\ref{con:eq0029}),(\ref{con:eq030}) are shown in Figure \ref{fig:fig3} (see middle and right panel). Generally, when $a$ takes small values, Equation (\ref{con:eq030}) can provide rather good values for $e_{max}$, but the expected time $T_{evol}$ given by Equation (\ref{con:eq0029}) is a little less than the actual time required to reach the maximum eccentricity. When $a$ takes large values, the quadrupole term effect becomes significant, and hence Equations (\ref{con:eq0029}),(\ref{con:eq030}) derived in dipole approximation may seriously misestimate the actual values of the maximum eccentricity and the evolution time. In addition, we have run some cases with different values of $m_d$, and the results show that the value of $e_{max}$ does not depend on $m_d$ and $T_{evol} \propto 1/m_d$, as expected from Equations (\ref{con:eq030}),(\ref{con:eq0029}).

\subsubsection{Quadrupole approximation $(\epsilon=1)$} \label{subsubsec:case12}

In the quadrupole approximation, solving $\dot{G}=0$, we still have $\omega=0 ,\pi/2,\pi,3\pi/2$.

At $\omega=0,\pi$, Equation (\ref{con:eq013}) becomes
\begin{equation}
\dot{g}=\frac{k}{G}\left(\frac{a}{R}\right)\left\{\frac{2-e^2-e^2\sin^2i}{2\sin i}-k_{0}(1-e^2)\right\}
\end{equation}
where $k_{0}=3\pi a/4R$. Solving $\dot{g}=0$, we get
\begin{equation}
\sin i=\frac{-k_{0}(1-e^2)+\sqrt{k_{0}^2(1-e^2)^2+e^2(2-e^2)}}{e^2}
\label{con:eq032}
\end{equation}
Since $0\leq\sin i\leq1$, in order to make the above equation true, the following inequality must be satisfied:
\begin{equation}
k_{0}\ge1\ \ \text{or}\ \ \frac{a}{R}\ge\frac{4}{3\pi}
\label{con:eq033}
\end{equation}
Thus, when $k_0>1$ (or $a/R\gtrsim 0.42$), $\dot{g}=0$ has solutions at $\omega=0,\pi$, and the system has the equilibrium points at $\omega=0,\pi$ (for certain values of $J_z$ constrained by Equation (\ref{con:eq032})). However, when $k_0<1$ (or $a/R\lesssim 0.42$), $\dot{g}=0$ has no solution at $\omega=0,\pi$ and hence there are no equilibrium points at $\omega=0,\pi$.

At $\omega=\pi/2,3\pi/2$, Equation (\ref{con:eq013}) becomes
\begin{equation}
\begin{aligned}
\dot{g}=-\frac{k}{G}\left(\frac{a}{R}\right)\left\{3(1-e^2)\sin i-\frac{\cos^2i}{\sin i}\left(1+\frac{3}{2}e^2\right)+\frac{k_0}{2}\big[3(1-e^2)-5\cos^2i\big]\right\}\\
\end{aligned}
\label{con:eq036}
\end{equation}
\begin{figure}[t]
\centering
\includegraphics[scale=0.28]{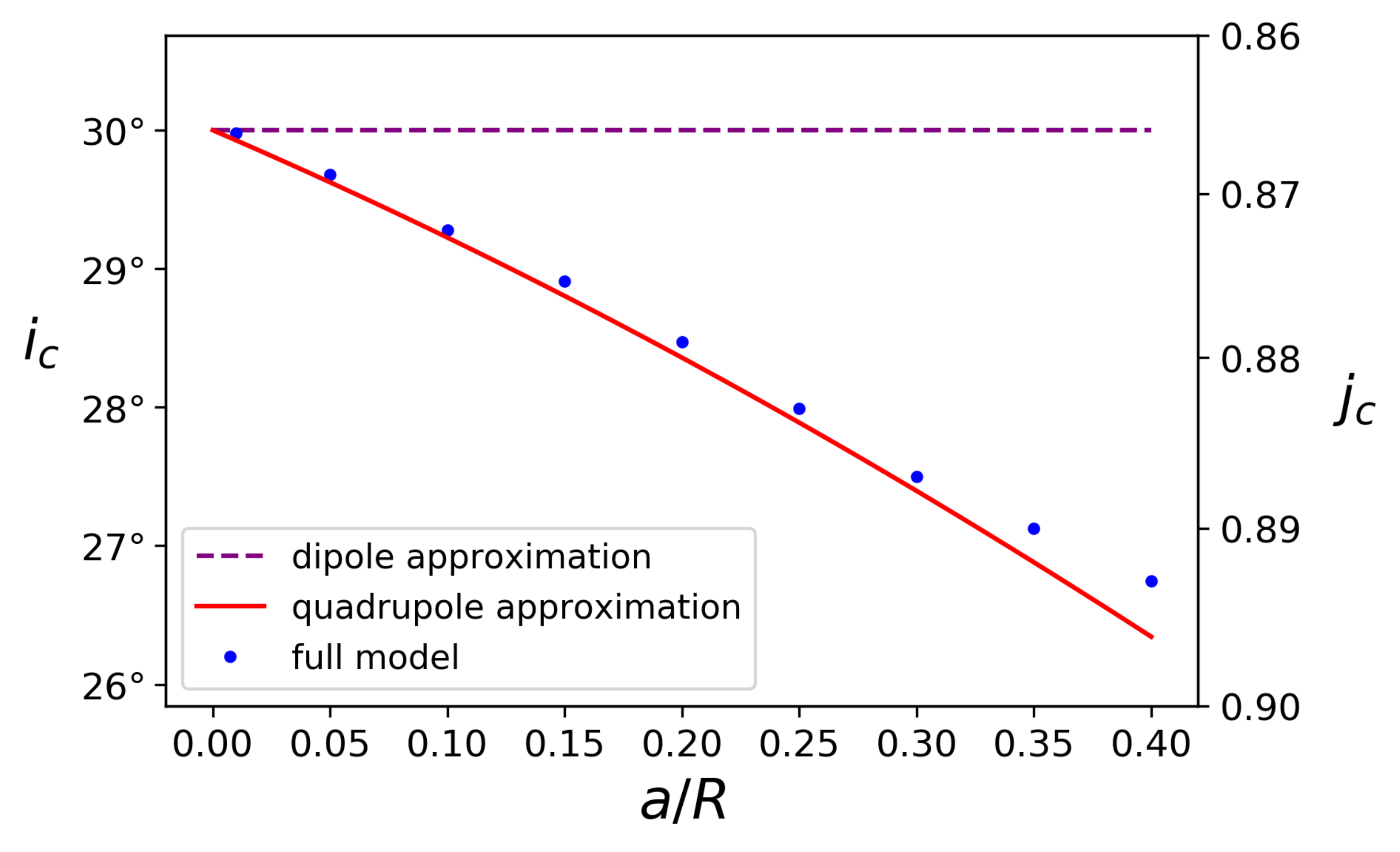}
\caption{The critical value $J_c$ and the critical inclination angle $i_c$ . The red line is computed in the quadrupole approximation, the purple line in the dipole approximation, and blue points in the full model (see Section \ref{sec:numerical}).}
\label{fig:fig05}
\end{figure}
\begin{figure}[t]

\gridline{\fig{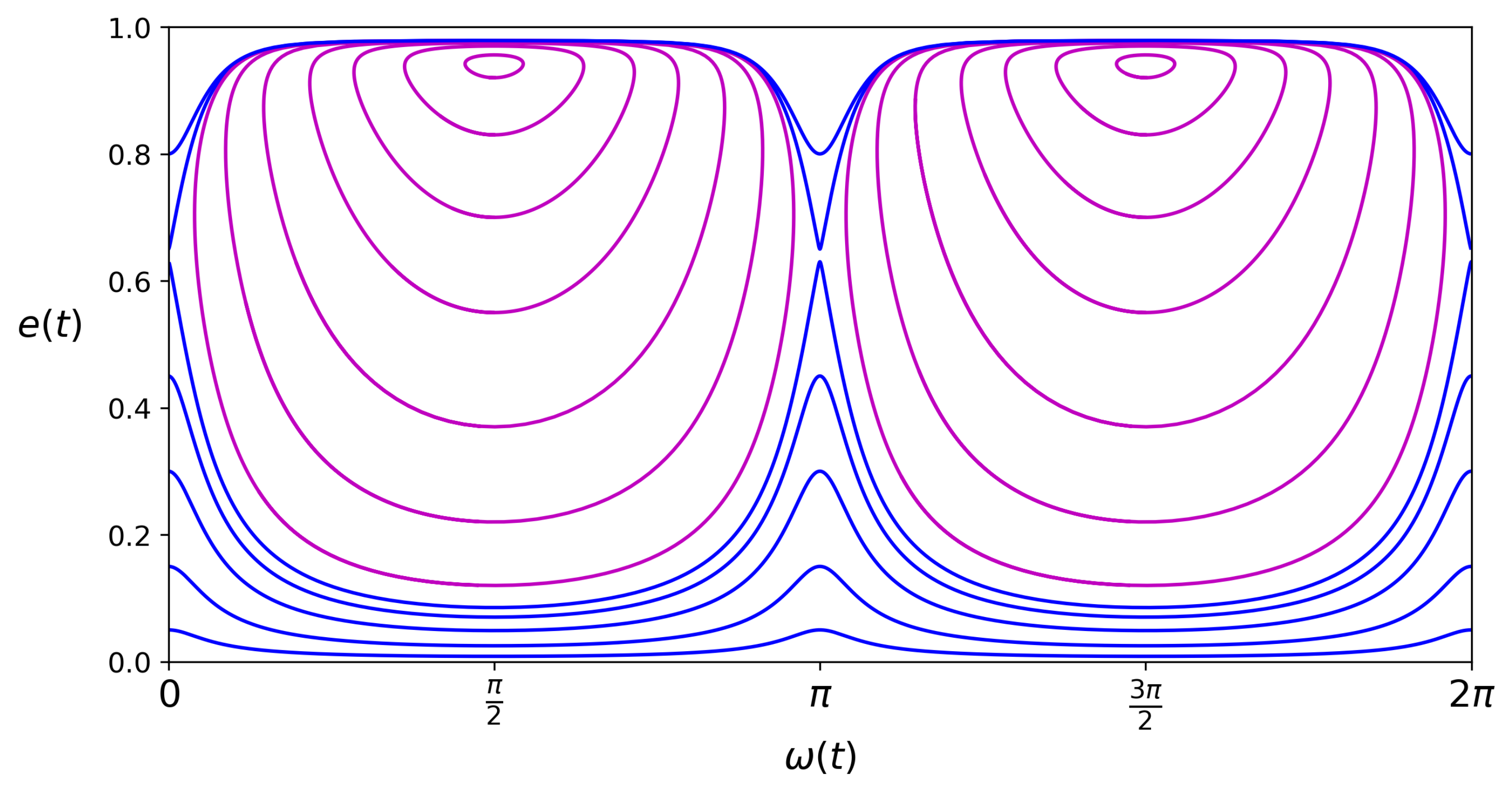}{0.42\textwidth}{(a)}
           \hspace{-3cm}
          \fig{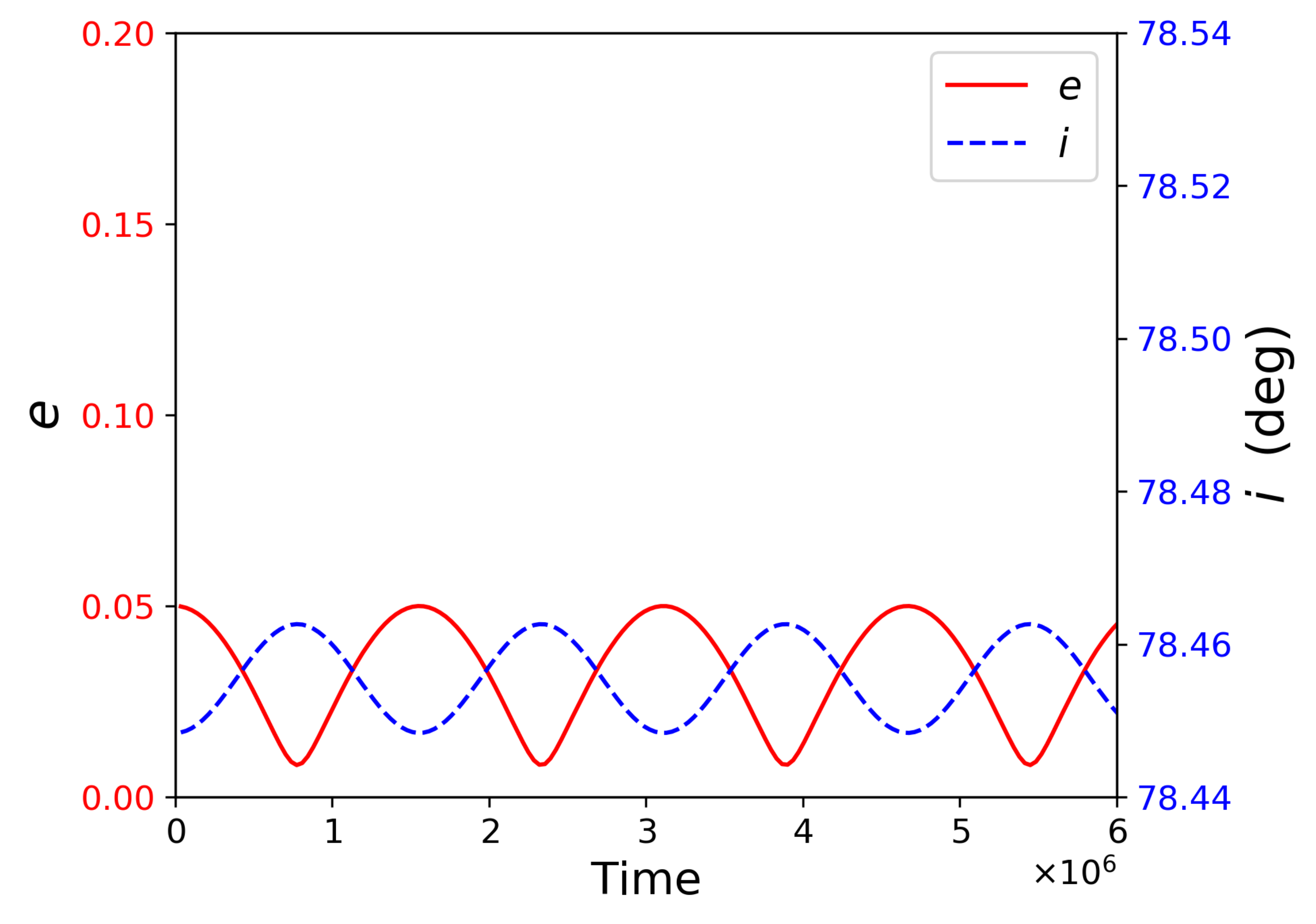}{0.3\textwidth}{(b)}
          }
\caption{(a): ($e,\omega$) phase space trajectories for $a=45$, $J_z=0.2$ in the quadrupole approximation. There are two unstable equilibrium points at $\omega=0,\pi$ and two stable equilibrium points at $\omega=\pi/2,3\pi/2$. (b): Periodic oscillations of $e$,$i$ over time for the orbit with $a=45$, $e_0=0.05$, $i_0=78.45^\circ$ (i.e. $J_z=0.2$). }
\label{fig:fig04}
\end{figure}
For a certain value of $k_0$ (or $a/R$), we can solve Equation (\ref{con:eq036}) numerically and then obtain the values of $J_z$ which make $\dot{g}=0$ has solutions at $\omega=\pi/2,3\pi/2$. Consequently, corresponding to these values of $J_z$, the system has equilibrium points only at $\omega=\pi/2,3\pi/2$ when $a/R<0.42$, and the orbits undergo the classical Lidov-Kozai resonance as shown in Figure \ref{fig:fig2}. Figure \ref{fig:fig05} provides the critical value $J_c$ and the corresponding critical angle $i_c$ for occurrence of the classical Lidov-Kozai resonance as a function of $a/R$ ranging from $0$ to $0.4$. If $J_z$ is smaller than the critical value $J_c$, the Lidov-Kozai resonance occurs. If $J_z$ is larger than the critical value $J_c$, there are no any secular resonances.

Comparison of the red line and the points in Figure \ref{fig:fig05} shows that the quadrupole approximation agrees very well with the full model for $a/R$ below 0.4 (in the full model, the potential of the uniform disk is neither approximated nor averaged). In the dipole approximation, $J_c$ and $i_c$ do not depend on the value of $a/R$. But in the quadrupole approximation, $J_c$ slightly increases from $\sqrt3/2\:(\approx0.866)$ to 0.896 (approximately) as $a/R$ increases from 0 to 0.4, meanwhile, $i_c$ drops from $30^\circ$ to $26.4^\circ$ (in the full model closer to $27^\circ$).

When $a/R>0.42$, $\dot{g}=0$ still has solutions at $\omega=\pi/2,3\pi/2$ for some values of $J_z$. However, as mentioned above, $\dot{g}=0$ may also have solutions at $\omega=0,\pi$. Consequently, the equilibrium points of the system may appear at $\omega=0,\pi/2,\pi,3\pi/2$. In this case, the phase space structure as well as the dynamical behaviors of the orbits are different from that of the classical Lidov-Kozai case of $a/R<0.42$. Figure \ref{fig:fig04}(a) shows a new ($e,\omega$) phase space structure with the equilibrium points at $\omega=0,\pi/2,\pi,3\pi/2$. One observes that the small eccentricities cannot be pumped to large values even at very high inclinations, and the corresponding inclination variations are also very small (see Figure \ref{fig:fig04}(b)). Hence, in this case the small eccentricity orbits can be maintained at a highly inclined configuration. This is a significant difference from the classical Lidov-Kozai case, in which the small eccentricities with high inclinations will be excited to large values and the inclined orbits are unstable.

In fact, when $a/R>0.42$, there are many other types of the Lidov-Kozai resonance. One of them has been shown in Figure \ref{fig:fig04}, and more types can be seen in Section \ref{sec:numerical}.

\subsection{Dynamics for the outer orbit} \label{subsec:case3}

\begin{figure}[h]
\centering
\includegraphics[scale=0.25]{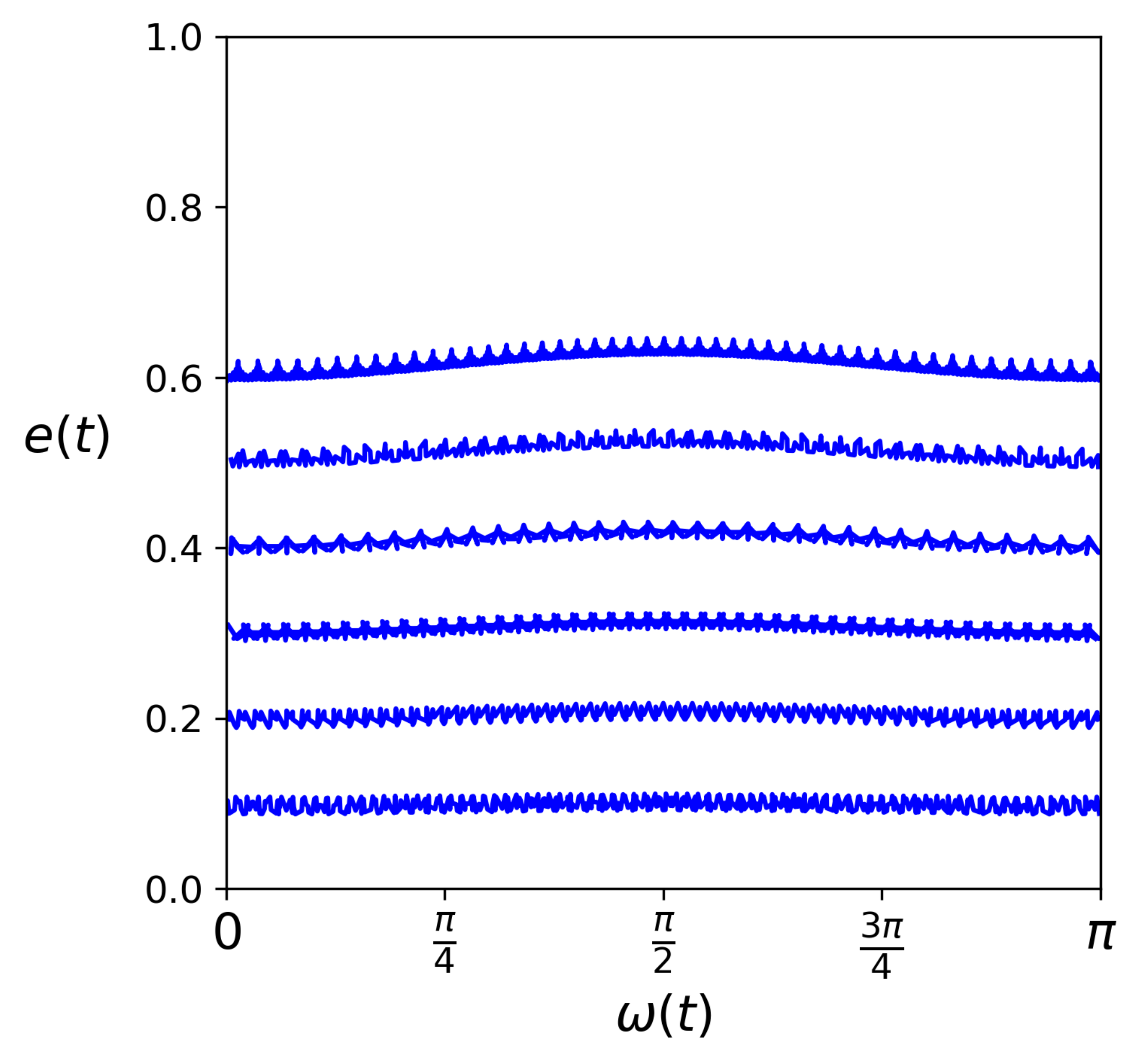}
\hspace{1.cm}
\includegraphics[scale=0.25]{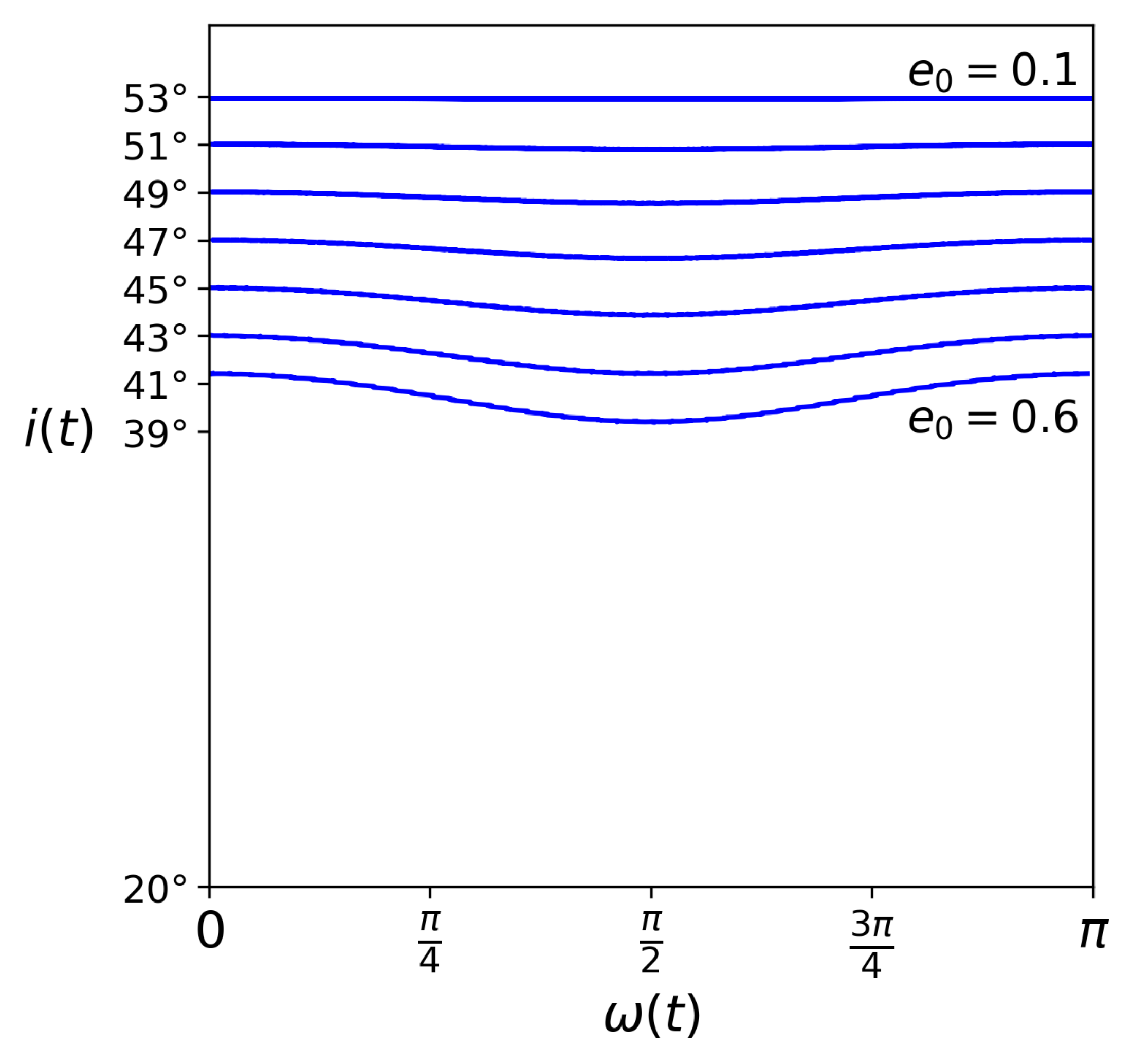}
\caption{$(e,\omega)$ and $(i,\omega)$ phase space trajectories for $a=300$, $J_z=0.6$ in the full model. In the phase spaces only the trajectories of the outer orbits are presented. In the right panel, the top/bottom trajectory corresponds to the orbit with $e_0=0.1$/$e_0=0.6$.}
\label{fig:fig4}
\end{figure}

In the outer orbit problem, the averaged Hamiltonian is
\begin{equation}
\overline{F}=\frac{\mu^2}{2L^2}+\frac{\mathcal{G}m_d}{R}\left\{\frac{R}{a}+\frac{1}{8}\left(\frac{R}{a}\right)^3\left(1-\frac{3}{2}\sin^2i\right)(1-e^2)^{-3/2}\right\}
\label{con:eq0039}
\end{equation}
Similarly, $L$,$H$ and $J_z$ remain constant. And
\begin{equation}
\begin{aligned}
&\dot{G}=\frac{\partial \overline{F}}{\partial g}\equiv0\\
&\dot{g}=-\frac{\partial \overline{F}}{\partial G}=\frac{k_1}{L^3G^4}\left(\frac{H^2}{G^2}-\frac{1}{5}\right)
\end{aligned}
\label{con:eq27}
\end{equation}
where $k_1=15\mathcal{G}m_d R^2/16$. Apparently, $G$ is constant, and hence $e$ as well as $i$ are also constant. This implies that variations of $e$ and $i$ are small in the full model. $\dot{g}$ remains as a non-zero constant (except for $H^2/G^2=1/5$ or $i=63.4^\circ$), which means the precession of $\omega$ from 0 to $2\pi$ is linear. Thus, the trajectories of the outer orbits are circulating in the phase space, and the outer orbits do not undergo secular resonances. Figure \ref{fig:fig4} shows the phase space trajectories for $a=300$, $J_z=0.6$ in the full model, and in Figure \ref{fig:fig4} only the trajectories of the outer orbits are presented. For these outer orbits, the variations of eccentricity and inclination are small.

We have run many cases of the outer orbit in the full model, and the results show that the variations of $e$ and $i$ are small for these outer orbits (even if $a/R$ is only a little greater than 1). Hence the outer orbits have strong stability.

\section{NUMERICAL STUDY} \label{sec:numerical}
In this section we perform our numerical study based on the full model. We introduce our full model first and show the validity of secular approximation within limits by comparisons with the full model. Then we focus more on the dynamics of the orbit of $a/R\sim1$ that is difficult to investigate by analytical methods.
\subsection{Full model} \label{subsec:case41}
Consider a massless test particle moving under the gravitational field of a central body and a uniform disk, the equation of motion for the massless particle is given by
\begin{equation}
\ddot{\boldsymbol{r}}=-\frac{\mu}{r^3}\boldsymbol{r}-\nabla V
\label{con:eq34}
\end{equation}
where $\boldsymbol{r}$ is the position vector of the particle. In the full model, the potential $V$ is neither approximated nor averaged. In order to easily compute the acceleration vector $\nabla V$, in the above equation of motion we adopt the closed form of the potential of uniform disk derived in \citet{lass1983gravitational}, instead of the integral form in Equation (\ref{con:eq2}). Accordingly, $\nabla V$ can also be written in closed form in terms of complete elliptic integrals, which can be computed precisely and easily. The detailed expressions of $\nabla V$ and its computational approaches can be found in \citet{krogh1982gravitational} and \citet{fukushima2010precise}. It needs to remark that the acceleration of the particle becomes infinite at the boundary of the uniform disk, hence we will terminate the calculations once the particle passes through the boundary. Equation (\ref{con:eq34}) is integrated using the Runge-Kutta-Fehlberg 7(8) integrator. In most cases, the integrator conserves the $z$-component of angular momentum and the energy of the orbit within the relative error of $10^{-4}$.

\begin{figure*}[h]
\gridline{\fig{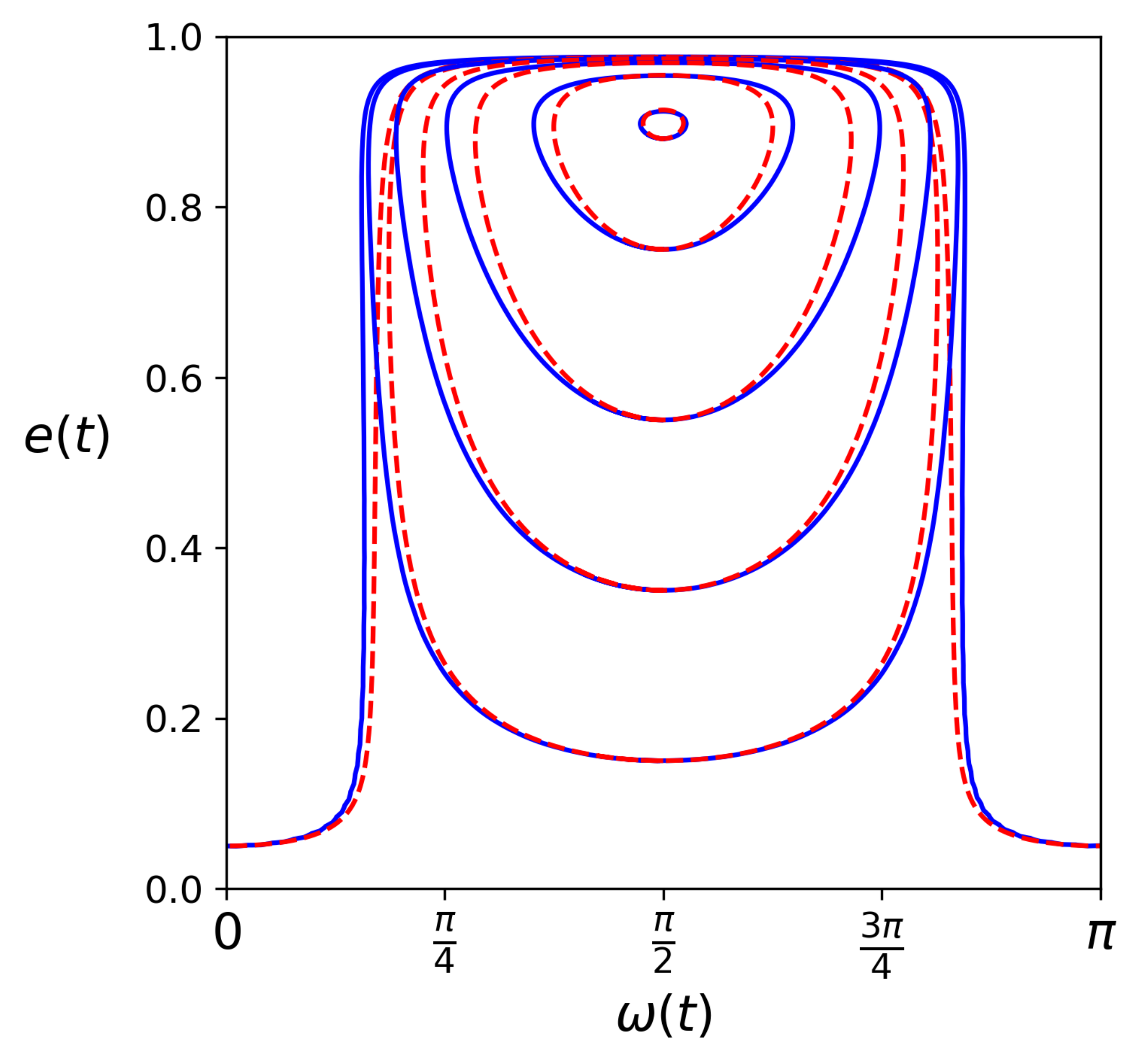}{0.25\textwidth}{(a)}
          \hspace{-5cm}
          \fig{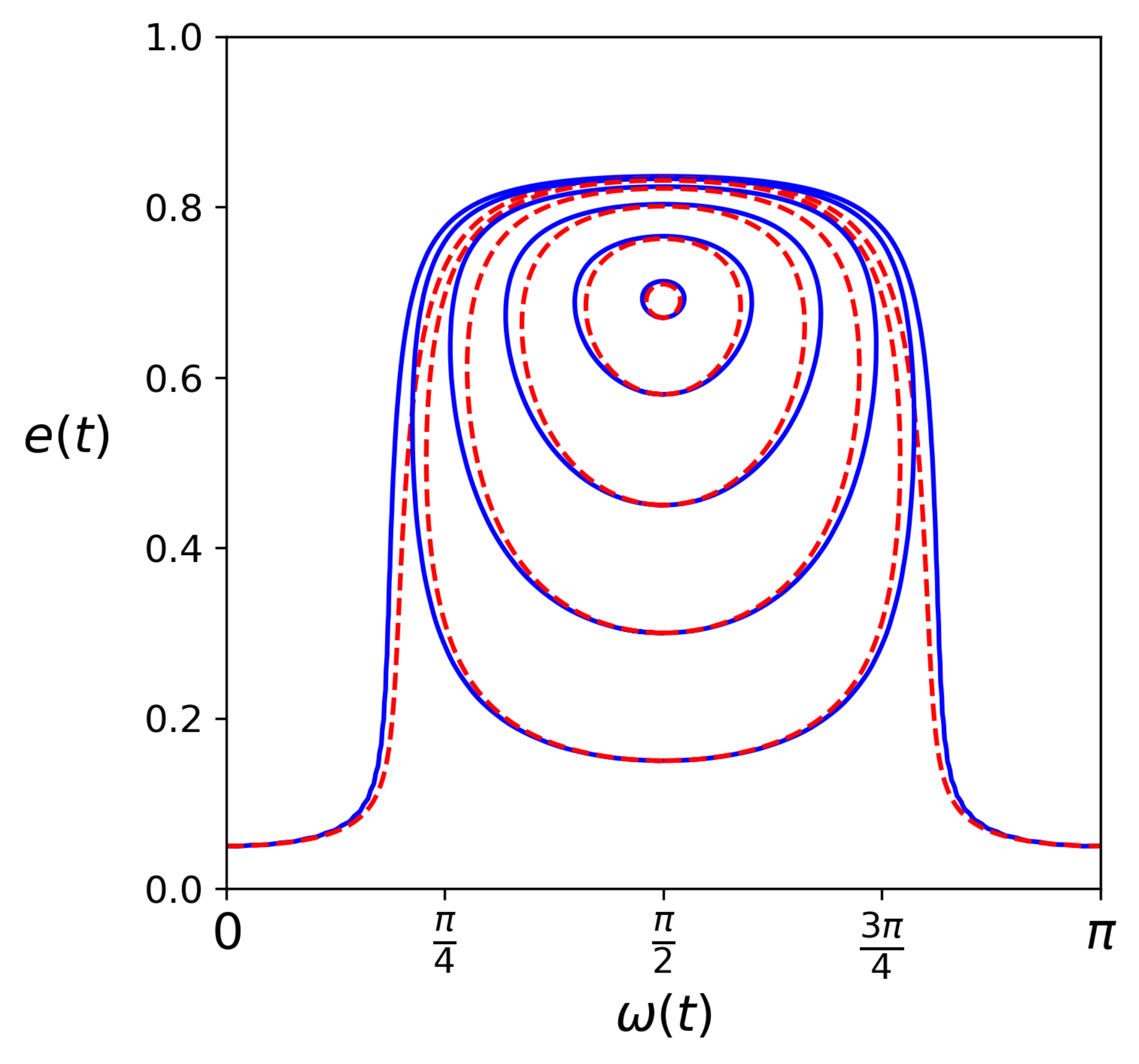}{0.25\textwidth}{(b)}
          }
\gridline{\fig{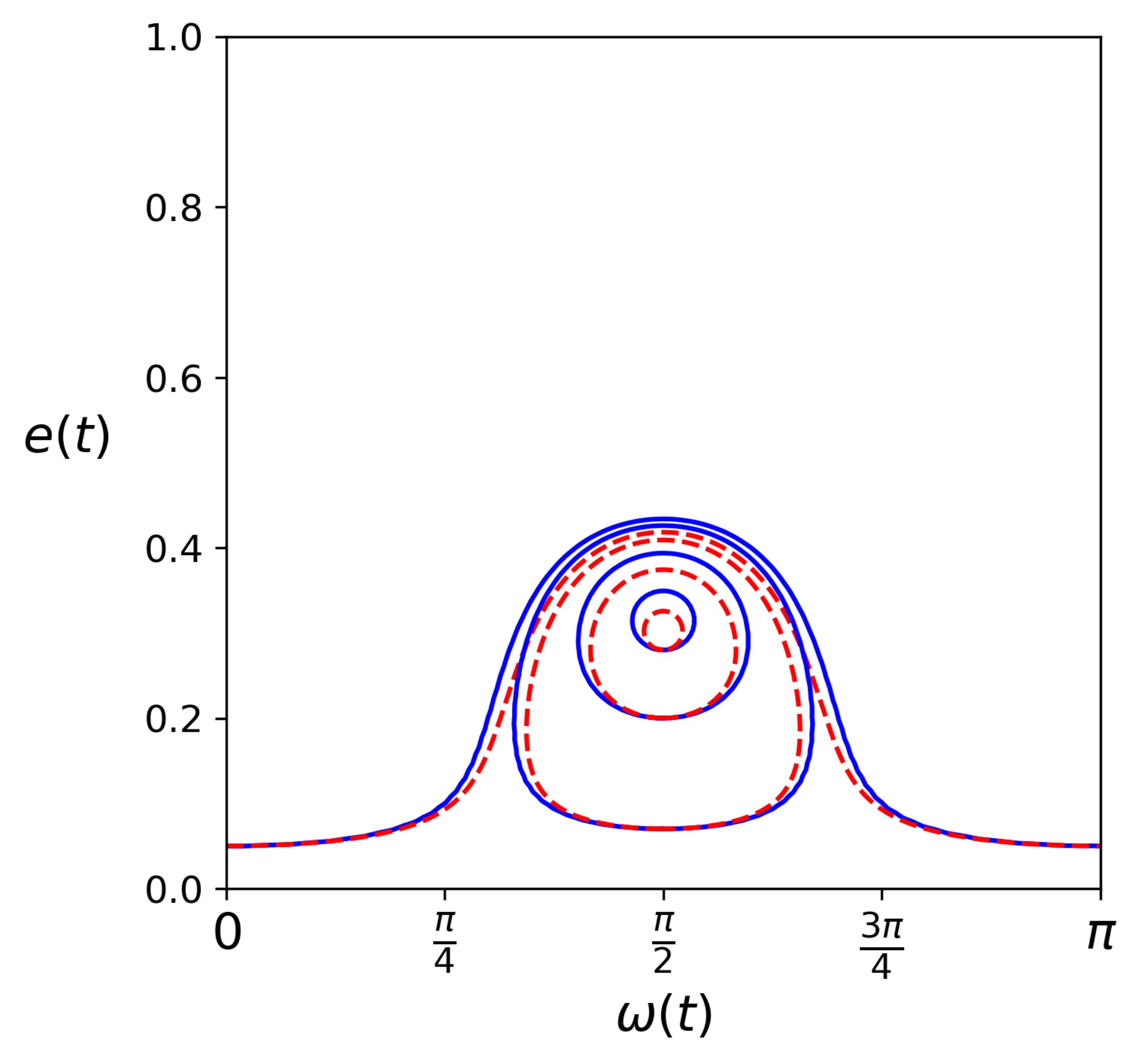}{0.25\textwidth}{(c)}
          \hspace{-5cm}
          \fig{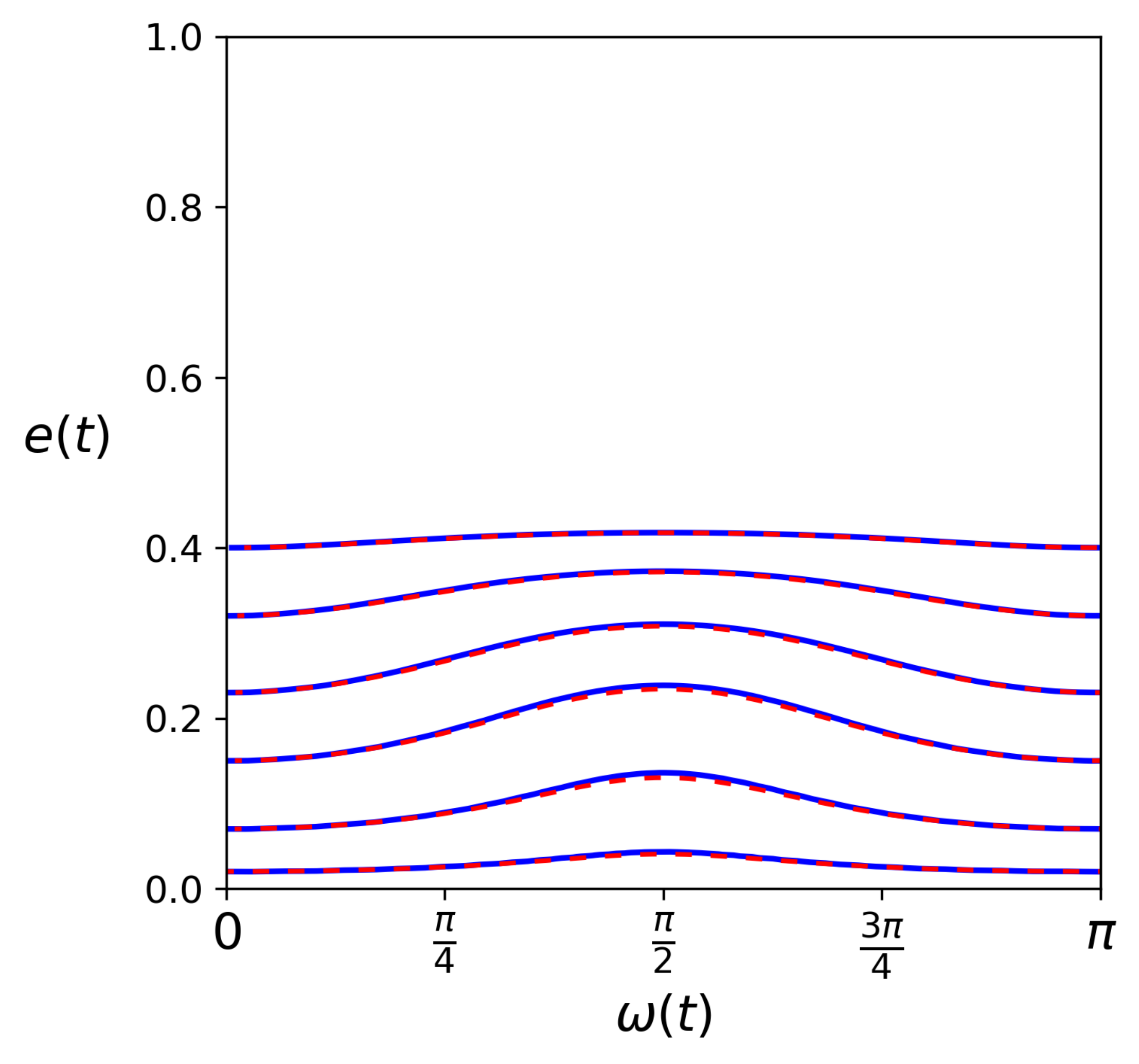}{0.25\textwidth}{(d)}
          }
\caption{Comparison between the full model and the dipole approximation on $(e,\omega)$ phase portraits for $a/R=0.1$. Where: (a) $J_z$=0.2;(b) $J_z$=0.5;(c) $J_z$=0.8;(d) $J_z$=0.9. In all panels, the blue solid lines describe the trajectories computed in full model, and the red dashed lines describe the trajectories in dipole approximation but starting from the same initial orbital elements as the trajectories in full model. }
\label{fig:fig5}
\end{figure*}

A comparison between full model and dipole approximation on $(e,\omega)$ phase portraits for $a/R=0.1$ is illustrated in Figure \ref{fig:fig5}. In these phase portraits, the equilibrium points and the trajectories given in the dipole approximation are almost identical with that of the full model. This indicates that the dipole approximation can sufficiently describe the dynamical behaviour of the orbit when $a/R$ takes a small value.

\begin{figure}[h]
\centering
\includegraphics[scale=0.23]{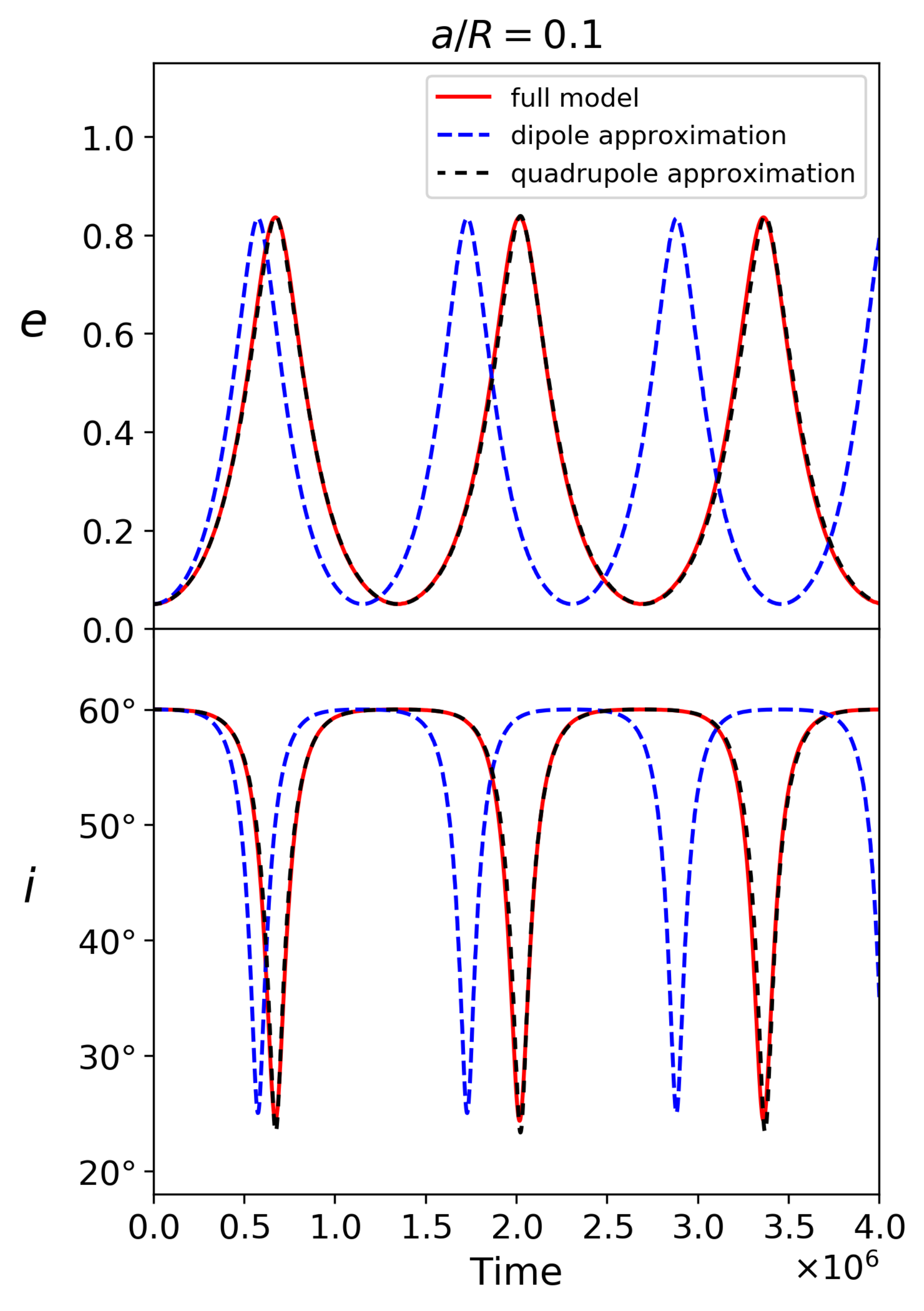}
\includegraphics[scale=0.23]{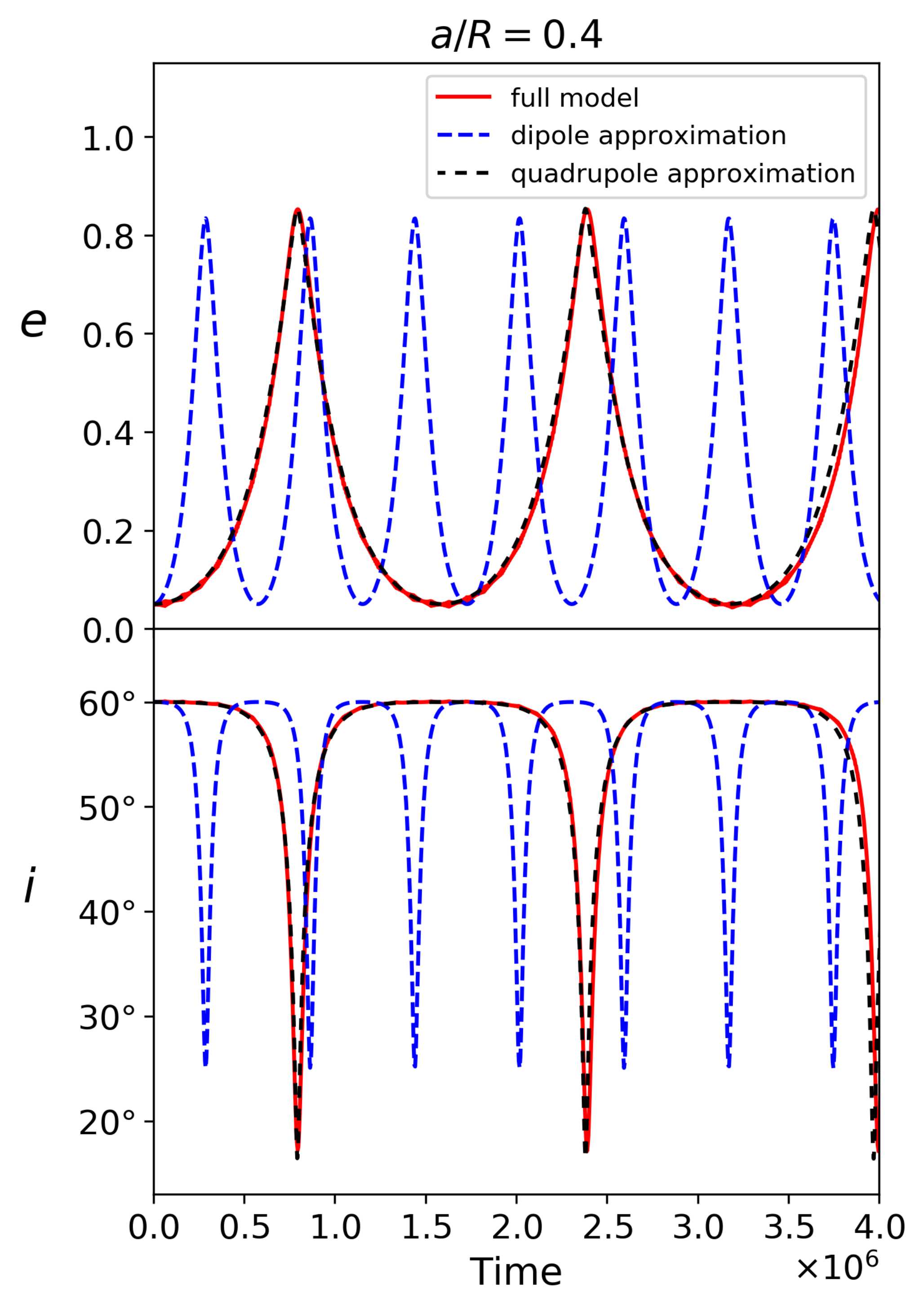}
\includegraphics[scale=0.23]{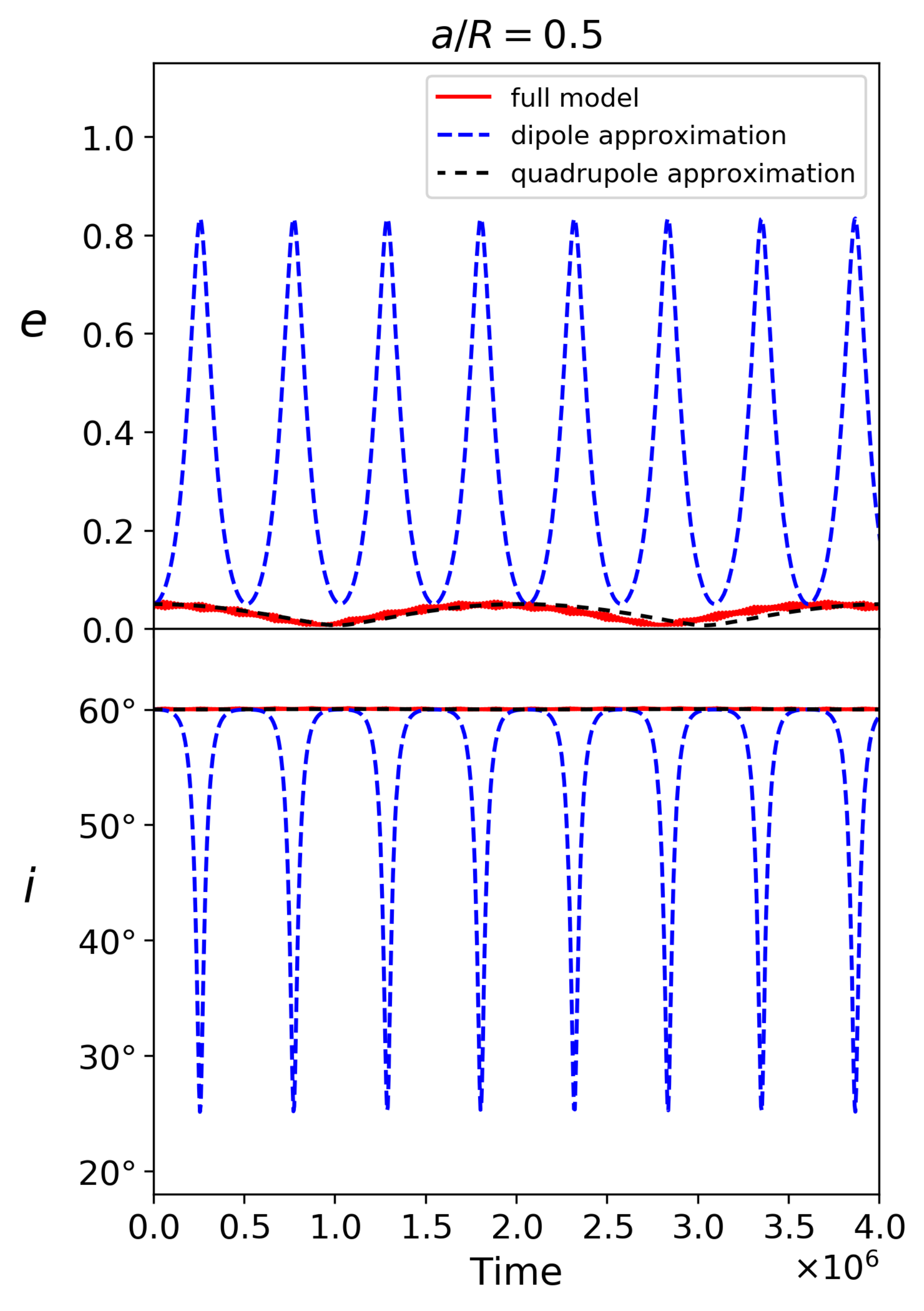}
\caption{The behaviours of eccentricity and inclination for the different values of $a/R$. We consider: $\mathcal{G}=1, M_{\star}=1, m_d=0.01,R=100$. The curves in all panels star with the same initial eccentricity and inclination: $e_0=0.05, i_0=60\degr$, but the left panel with the semimajor axis $a=10$, the middle panel: $a=40$, and the right panel: $a=50$. The red solid lines represent the curves computed in full model, the blue dash lines in dipole approximation, and the black dash lines in quadrupole approximation.}
\label{fig:fig6}
\end{figure}

For different values of $a/R$, a comparison between full model and dipole/quadrupole approximation on the behaviours of the eccentricity and inclination is shown in Figure \ref{fig:fig6}. One observes: The quadrupole approximation agrees very well with the full model even $a/R=0.5$, which suggests that the quadrupole approximation is still valid for large values of $a/R$.  The dipole approximation and the full model are in good overall agreement up to $a/R=0.4$, except for the oscillation period in the case $a/R=0.4$. In the case $a/R=0.5$, the dipole approximation is significantly inconsistent with the full model. The eccentricity and inclination variations in the dipole approximation are very large, and the initial small eccentricity is still excited to a large value due to the dipole-level Lidov-Kozai effect which does not depend on the value of $a/R$. However, as illustrated in Section \ref{subsubsec:case12}, when $a/R>0.42$ the excitation of the small eccentricity could be ``suppressed" induced by the quadrupole effect. As a result, in the full model/quadrupole approximation the eccentricity and inclination variations are small for $a/R=0.5$.
\subsection{Dynamics for the orbit of $a/R\sim1$} \label{subsec:case42}
\begin{figure*}[htbp]
\centering
\gridline{\fig{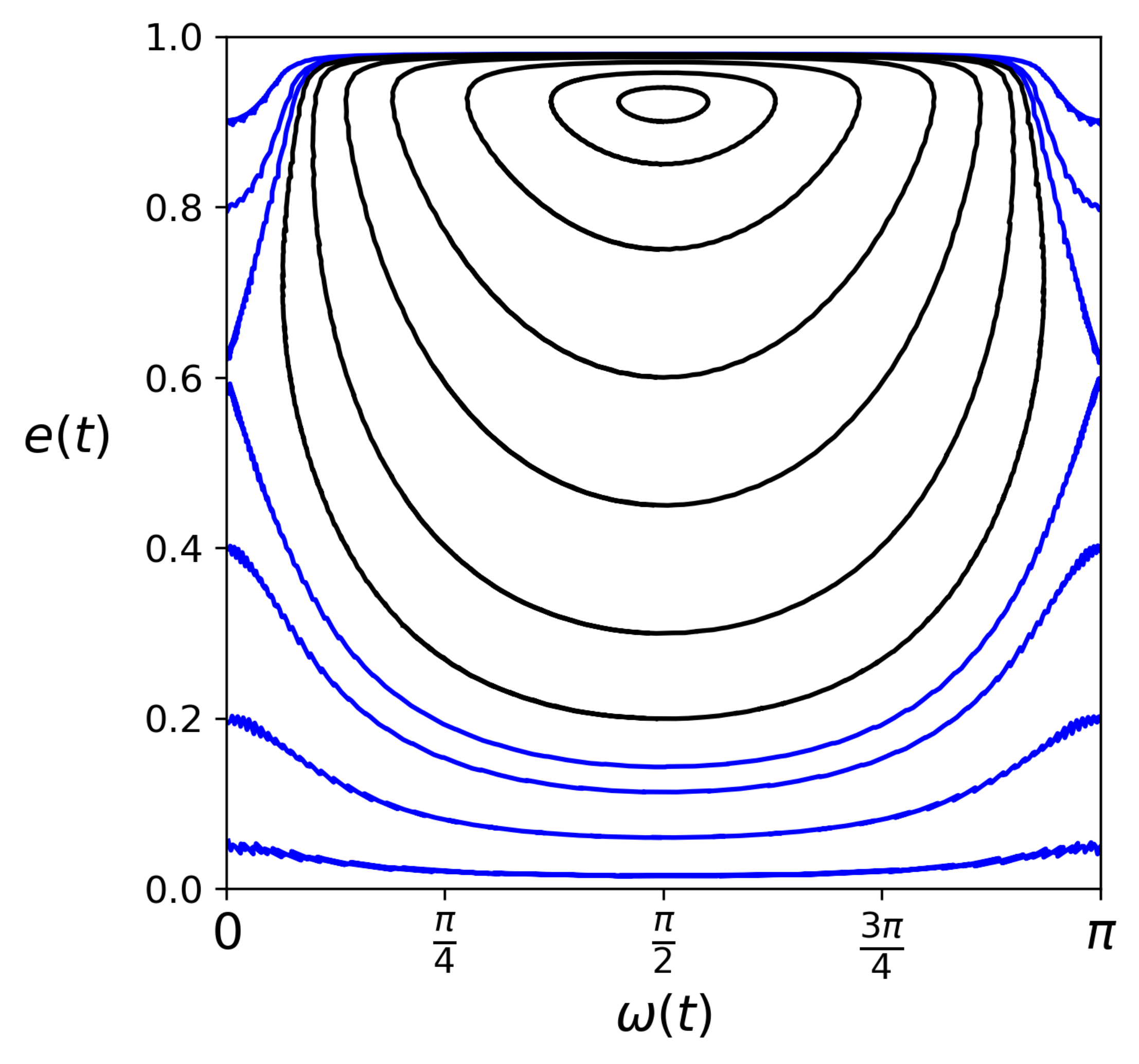}{0.27\textwidth}{(a)}
          \fig{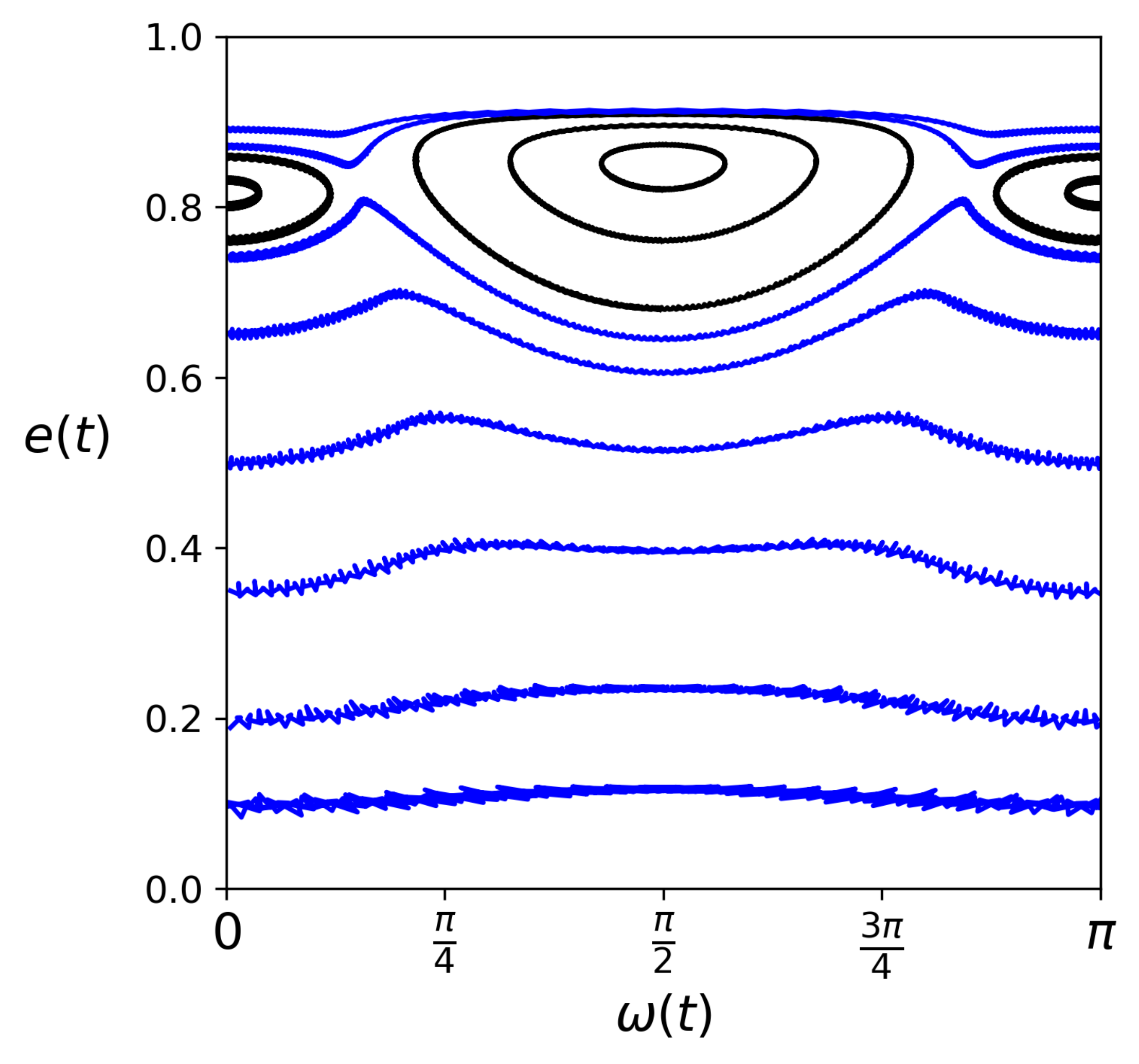}{0.27\textwidth}{(b)}
          \fig{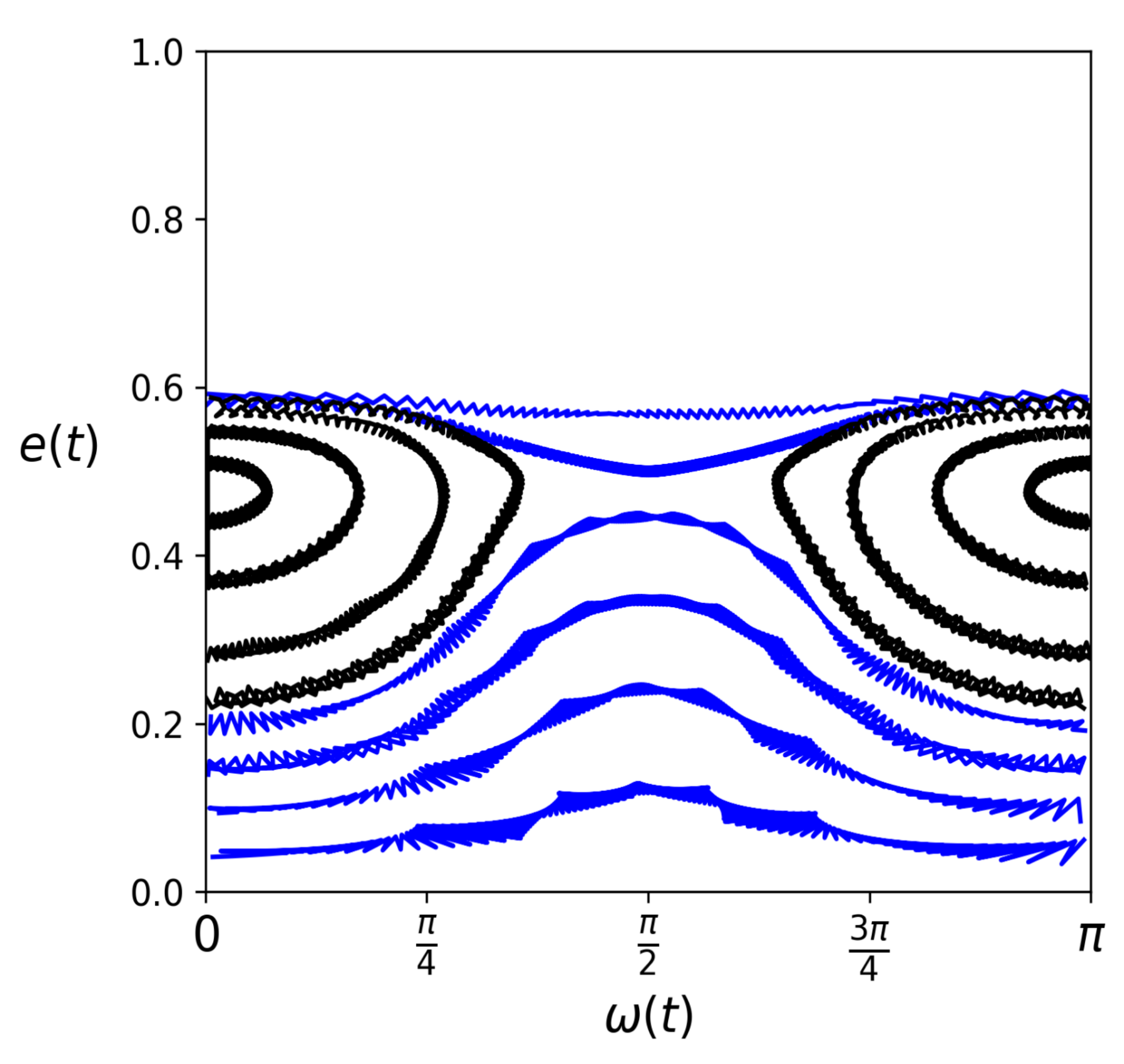}{0.27\textwidth}{(c)}}
\gridline{\fig{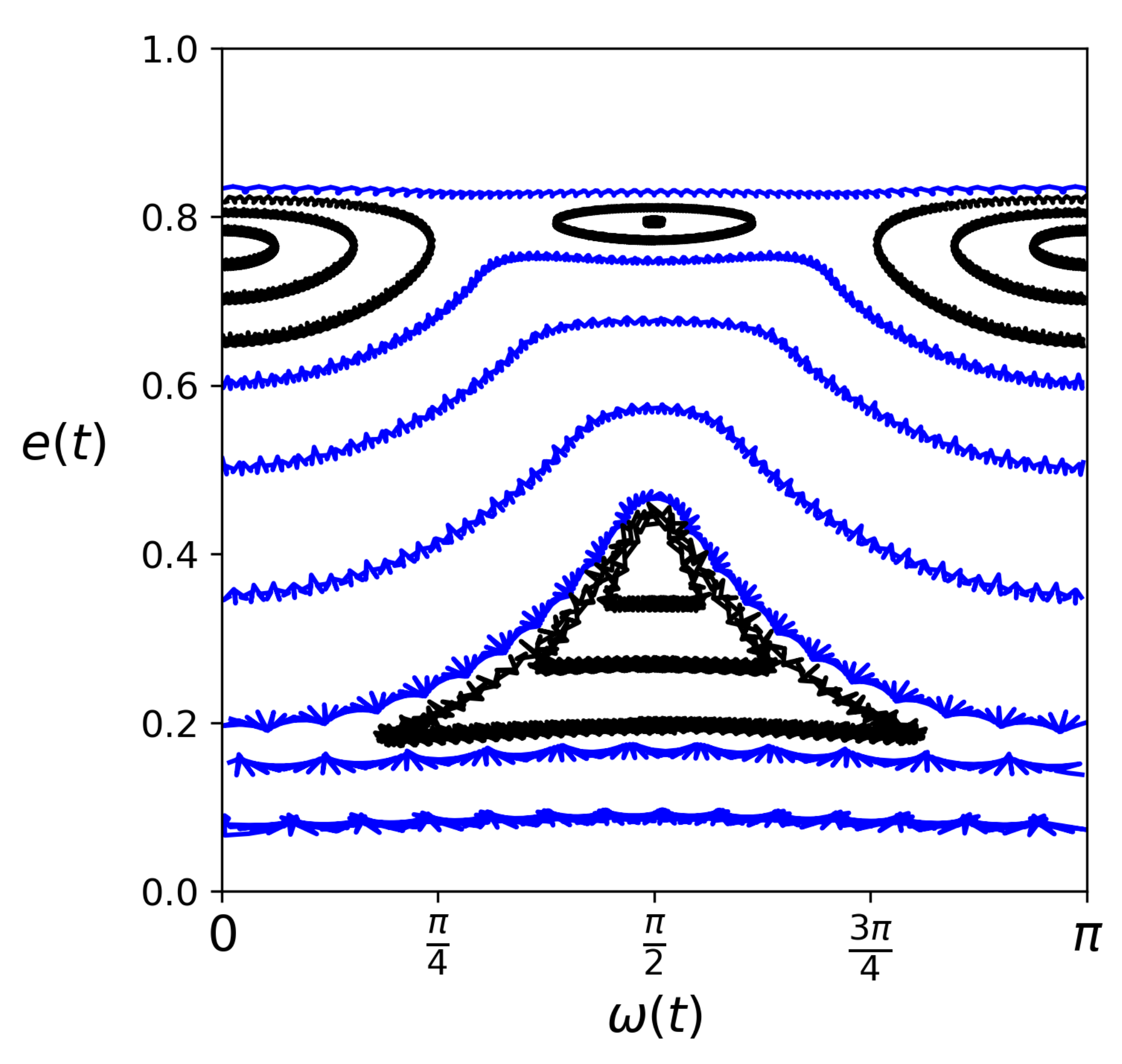}{0.27\textwidth}{(d)}
         \fig{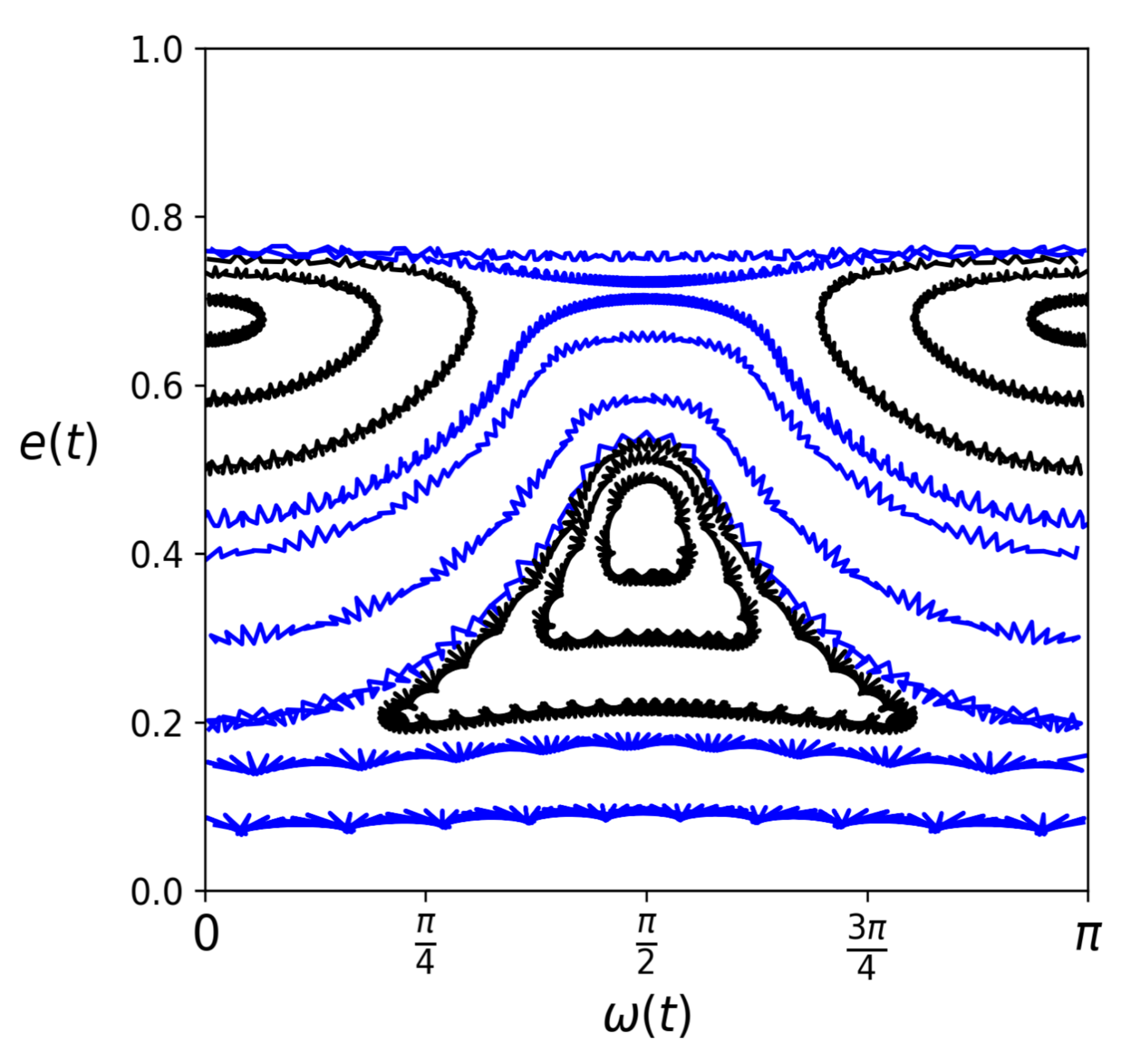}{0.27\textwidth}{(e)}
          \fig{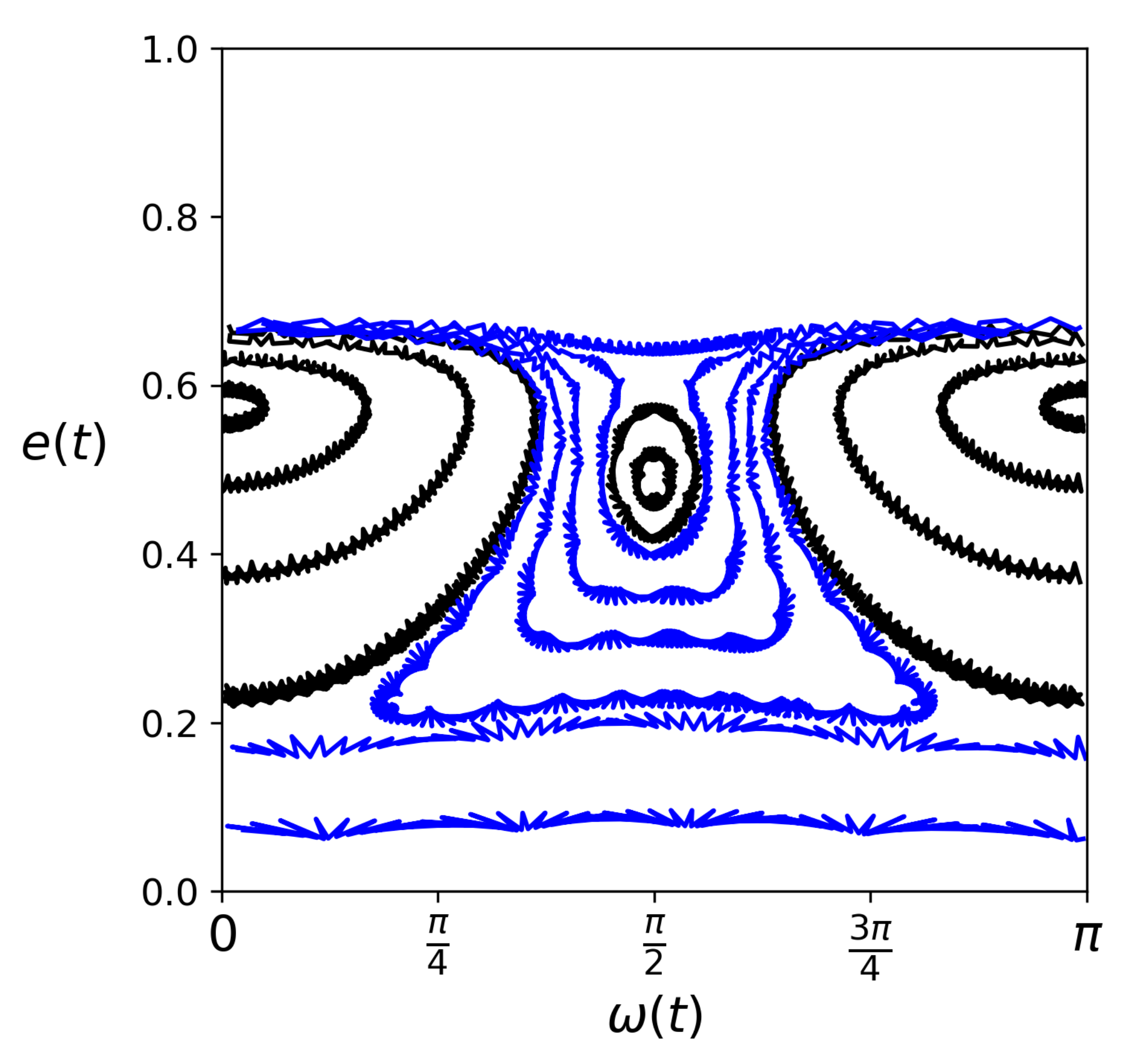}{0.27\textwidth}{(f)}}
\gridline{\fig{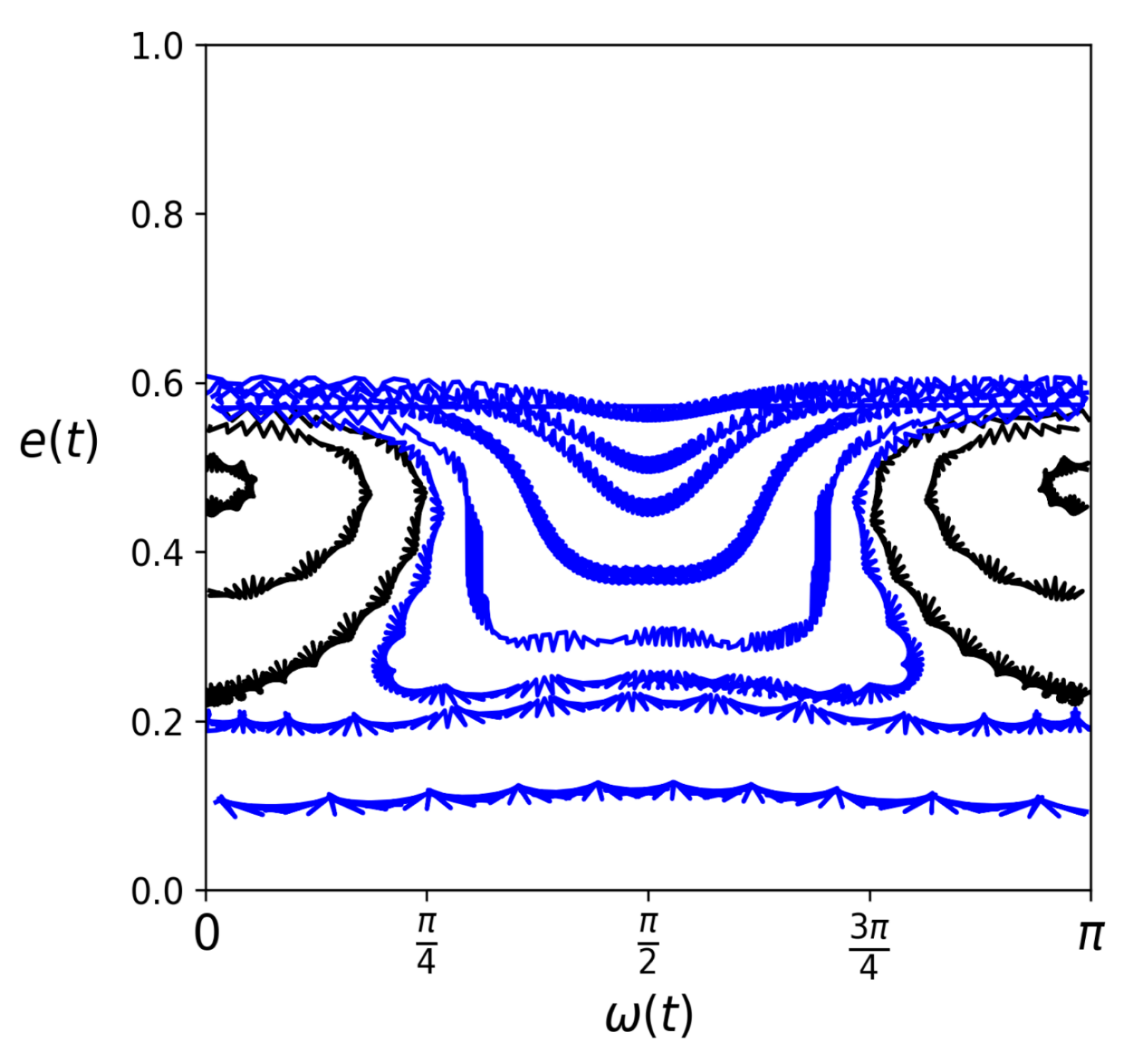}{0.27\textwidth}{(g)}}
\caption{$(e,\omega)$ phase space portraits for the case of $a/R\sim1$ in the full model. Where: (a) $a=50, J_z=0.2$;(b) $a=80, J_z=0.4$;(c) $a=90, J_z=0.8$;(d) $a=120, J_z=0.55$;(e) $a=120, J_z=0.65$;(f) $a=120, J_z=0.74$;(g) $a=120,J_z=0.8$ (the disk radius $R$ is 100). The black lines correspond to librating trajectories, and the blue lines to the circulating trajectories.  The equilibrium points appear at $\omega=0,\pi/2,\pi$, and other positions, for example, at $\omega=\pi/6, 5\pi/6$ approximately (see panel(b)).}
\label{fig:fig7}
\end{figure*}
We have analytically surveyed the dynamics of the inner orbit and outer orbit under the secular perturbation of a uniform disk. However, in the previous analysis, we only considered the limiting case of $a/R$ taking small (or large) values. For the case of $a/R\sim1$, octupole and higher order terms become nonnegligible and would need to be taken into account. On the other hand, once the orbits cross the sphere surface $r=R$, our Equations (\ref{con:eq8})(\ref{con:eq9}) break down. Thus, it is rather difficult to study the dynamical behaviour of the orbit of $a/R\sim1$ through analytical methods, we have to resort to numerical methods.

We have carried out massive numerical calculations for the case $a/R\sim1$ using the full model. And the results show that the orbits could undergo many more complicated Lidov-Kozai resonances which are different from the classical type (depicted in Figure \ref{fig:fig2}). Figure \ref{fig:fig7} illustrates seven resonant phase space structures, each of which corresponds to a special Lidov-Kozai resonance type. These resonances shown in Figure \ref{fig:fig7} (a),(b),...,(g) will be called Type a,b,...,g, respectively. One observes that these resonances all have equilibrium points at $\omega=0,\pi$. In Type d and Type e there are two equilibrium points at $\omega=\pi/2$, and in Type g the equilibrium point disappears at $\omega=\pi/2$. In particular, in Type b,d the equilibrium points appear at other values of $\omega$ (besides $\omega=0,\pi/2,\pi$) and these values are not fixed. We have demonstrated that the quadrupole term effect can lead to the equilibrium points at $\omega=0,\pi$ in Type a. Consequently, it is plausible to suppose that the appearances of the equilibrium points in the other types are also attributed to the high-order term effects.

\begin{figure}[htbp]
\centering
\includegraphics[scale=0.5]{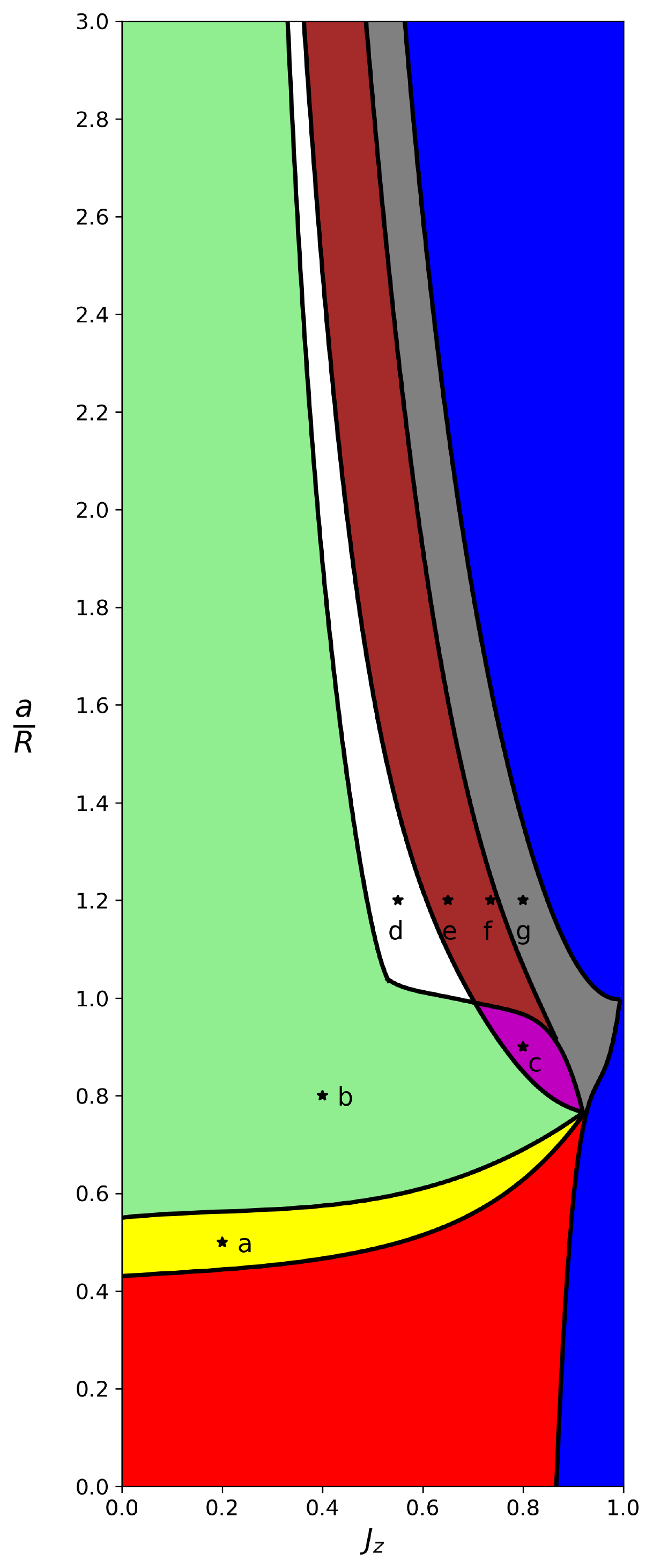}
\caption{Distribution of all resonance types in the parameter space $a/R\times J_z$, for the secular problem of the uniform disk perturbation. The red region represents the distribution of the classical Lidov-Kozai resonance as shown in Figure \ref{fig:fig2}(a). The blue region represents the distribution of the non-resonance type shown in Figure \ref{fig:fig2}(d), and the other regions, yellow: Type a; green: Type b; purple: Type c; white: Type d; brown: Type e and Type f; gray: Type g. The points corresponding to the phase space portraits shown in Figure \ref{fig:fig7} are highlighted in black stars.}
\label{fig:fig8}
\end{figure}

In fact, which type of resonance the orbit will undergo under the uniform disk perturbation is determined by the parameters $a/R$ and $J_z$. We obtain the distributions of the all resonance types in the parameter space $a/R\times J_z=[0,3]\times[0,1]$ through global numerical calculations (see Figure \ref{fig:fig8}). In our runs, the parameter space in $a/R$ is covered in steps of 0.01 and $J_z$ in steps of 0.01. In Figure \ref{fig:fig8}, the red region gives the distribution of the classical type (as shown in Figure \ref{fig:fig2}(a)), the yellow region gives the distribution of Type a, and green: Type b; purple: Type c; white: Type d; gray: Type g; brown: Type e,f. The blue region is non-resonance region where the Lidov-Kozai resonance does not occur (see Figure \ref{fig:fig2}(d)). As Type e and Type f both have the same number of the stable and unstable equilibrium points, we merge them into the same distribution for simplicity. The differences between ($e,\omega$) phase space portraits caused by varying $a/R$ or $J_z$ in the same distribution region are only the size of libration island, the oscillation amplitudes of $e$ and $\omega$, and the positions of the equilibrium points.

As mentioned in the quadrupole approximation, when $a/R>0.42$ there are such Lidov-Kozai resonances in which the system has equilibrium points at $\omega=0,\pi$. The numerical results show the critical value of $a/R$ in the full model is about 0.43, which corresponds to the value of $a/R$ of the boundary point between the red region and yellow region at $J_z=0$ in Figure \ref{fig:fig8}. The analytical value of 0.42 is rather close to the numerical value of 0.43.

In the case of $a/R\sim1$, the most main resonance is Type b corresponding to the green region in Figure \ref{fig:fig8}. Liking in Type a, the small eccentricities in Type b cannot grow to large values even at high inclinations, and hence the corresponding inclinations also cannot drop to very low values (as $\sqrt{1-e^2}\cos i$ remains constant). As a result, in Type b the small eccentricity orbits can be maintained at a highly inclined state. Overall, in the case of $a/R\sim1$, the libration islands are a little small in comparison with the case of $a/R\ll 1$, the orbital resonances are not very dramatic, and the variations of the eccentricity as well as the inclination are relatively moderate.

\begin{figure}[htbp]
\centering
\includegraphics[scale=0.27]{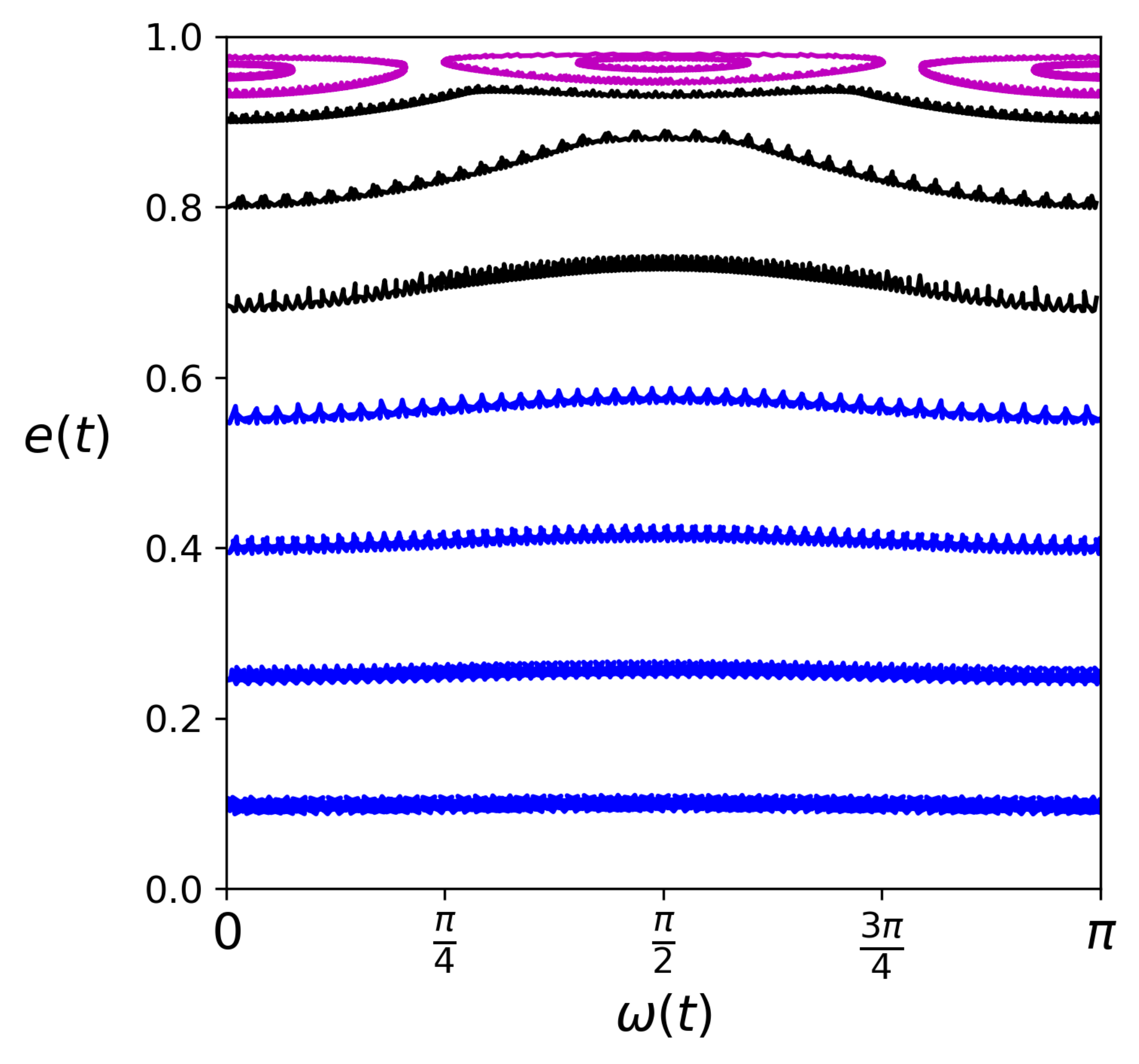}
\caption{Phase space trajectories for $a=300$,$J_z=0.2$ in the full model. The trajectories in libration islands are depicted in magenta lines. The trajectories of the outer orbits are depicted in blue lines and they are circulating.}
\label{fig:fig10}
\end{figure}

In the region of $a/R>1$, the resonances shown in Figure \ref{fig:fig7} gradually fade away as $a/R$ increases, and the blue non-resonance region becomes larger and larger (see Figure \ref{fig:fig8}), more and more orbits no longer undergo secular resonances. Although when $a/R$ is large, such as $a/R=3$, the orbital resonances still exist for some values of $J_z$, the libration islands are rather small in these resonances and appear at high eccentricities (see Figure \ref{fig:fig10}). The most low eccentricity ($e\lesssim 0.6$) orbits (actually the outer orbits) do not undergo secular resonance, their trajectories are circulating from $0$ to $2\pi$ and the eccentricity variations are small.
\section{CONCLUSION AND DISCUSSION \label{sec:discussion}}

In this paper, we have studied the secular behaviour of a particle's orbit under the gravitational perturbation from a uniform disk. By averaging the multipole expansion of the disturbing potential over the orbit, we develop the secular approximation for the secular problem. We can analytically derive some properties of the system based on the secular approximation.

For the inner orbit problem, we first consider the dipole level of the secular approximation, i.e., the dipole approximation. We demonstrated that when the Kozai integral $J_z\leq \sqrt{3}/2$, the Lidov-Kozai resonance occurs for the inner orbits and the system has two equilibrium points at $\omega=\pi/2,3\pi/2$. The critical value $\sqrt{3}/2$ corresponds to the critical inclination $30^\circ$ above which the eccentricity and inclination variations of the orbits are usually large. The maximum eccentricity $e_{max}$ reached by the eccentricity depends only on the initial inclination $i_0$, and $e_{max}$ increases as $i_0$ increases (for the prograde orbits). For the very large value of $i_0$, the eccentricity can be excited to near 1 due to the Lidov-Kozai effect. The oscillation period or evolution time $T_{evol}\propto 1/m_d$. When $a/R\ll1$, the dipole approximation agrees well with the full model. When $a/R$ takes larger values, the dipole approximation is inadequate to describe the behaviour of the system, and hence we need to take into account the quadrupole approximation.

In the quadrupole approximation, the critical value of $J_c$ for the occurrence of the Lidov-Kozai resonance slightly increases from $\sqrt{3}/2$($\approx 0.866$) to 0.896 as $a/R$ increases from 0 to 0.4 due to the quadrupole effect, and the corresponding critical inclination $i_c$ drops from $30^\circ$ to about $26.4^\circ$ (the value in the full model is closer to $27^\circ$). When $a/R>0.42$, besides $\omega=\pi/2,3\pi/2$, the equilibrium points of the system could also appear at $\omega=0,\pi$, which leads to the behaviours of the orbits different from that in the classical Lidov-Kozai resonance of $a/R<0.42$.

For the outer orbit problem, we find that the outer orbits do not undergo the Lidov-Kozai resonance under the secular perturbation of the uniform disk. The variations of the eccentricity and inclination for the outer orbits are small, and the outer orbits have strong stability.

We investigate the secular dynamics of the orbits with $a/R\sim1$ through the numerical methods. We find that there are many more complicated Lidov-Kozai resonance types in which the equilibrium points of the system appear at $\omega=0,\pi/2,\pi,3\pi/2$, even other values of $\omega$.  The eccentricity (as well as inclination) oscillations in these types are relatively moderate on the whole. In particular, in some resonance types the highly inclined orbits are stable.

We also find that the multipole expansion of the potential due to a Kuzmin disk \citep{kuzmin1956model} is similar to that of the uniform disk in form (see Appendix \ref{sec:A3}). And the difference between them is only in the numerical factor which does not affect the results of qualitative analysis. This implies that the orbit under the secular perturbation of the Kuzmin disk has the similar dynamical behavior with that under the uniform disk. As a result, for the orbit located at the central region of the Kuzmin disk, the Lidov-Kozai effect kicks in for the orbit if its inclination is larger than the critical value $30^{\circ}$ (in the dipole approximation). In addition, \cite{terquem2010eccentricity} mentioned that under the perturbation of the disk with the decreasing surface density $\sigma(\rho)\propto \rho^{-1/2}$, inner radius $R_i=1$AU and outer radius $R_o=100$AU (case A in the paper), the critical inclination $i_c$ for the eccentricity growth is also about $30^{\circ}$.

For the problem of the annulus disk with a non-zero inner radius $R_i$, \cite{terquem2010eccentricity} found that the Lidov-Kozai effect of $i_c=39.2^{\circ}$ will occur for the orbit that is well inside the inner cavity of the annulus (i.e. $a\ll R_i$). We have checked the behaviour of the orbit under the perturbation of the unform annulus with a small radius ratio $R_i/R_o$ (below 0.5). When $a\ll R_i$, the orbit is indeed subject to the Lidov-Kozai effect of $i_c=39.2^{\circ}$. For the orbit crossing the annulus, we find that the behaviour of the orbit is similar to that in the uniform disk case. In fact, since the radius ratio $R_i/R_o$ is small, to a certain extent the cumulative effect of the perturbation of the annulus is equivalent to that of the disk with $R_i=0$. Thus, the annulus perturbation will not significantly change the previous dynamical behavior of the orbit.

\acknowledgments
We thank the anonymous referee for helpful comments and suggestions improving the paper. This work is supported by the Chinese Academy of Sciences, the National Natural Science Foundation of China (NSFC) (Nos. 11673053,11673049), and The Youth Innovation Promotion Association of Chiness Academy of Sciences (2019265).

%




\appendix
\section{MULTIPOLE EXPANSION OF THE GRAVITATIONAL POTENTIAL OF UNIFORM DISK\label{sec:A1}}

1. When $r/R < 1$, it is difficult to expand the potential function in Equation (\ref{con:eq2}) directly into a power series of $r/R$. For this we need to perform some operations on Equation (\ref{con:eq2}) first. By integrating with respect to $\rho$, the potential function can be decomposed into:
\begin{equation}
\begin{aligned}
V&=-2\mathcal{G}\sigma\int_{0}^{R}\int_{0}^{\pi}\frac{\rho d\rho d\theta}{\sqrt{\rho^2+r^2-2\rho r\cos\theta\cos\varphi}}\\
&=-2\mathcal{G}\sigma\int_{0}^{\pi}\left\{\sqrt{R^2+r^2-2rR\cos\varphi\cos\theta}-r\right.\\
&\:\:\:\:\:\left.+r\cos\varphi\cos\theta\int_{0}^{R}\frac{d\rho}{\sqrt{\rho^2+r^2-2\rho r\cos\theta\cos\varphi}}\right\}d\theta\\
&\triangleq-2\mathcal{G}\sigma\left[-\pi r+I_1+r\cos\varphi(I_2-I_3)\right]
\end{aligned}
\label{con:A1}
\end{equation}
where
\begin{equation}
I_1=\int_{0}^{\pi}\sqrt{R^2+r^2-2rR\cos\varphi\cos\theta}\:d\theta
\end{equation}
\begin{equation}
I_2=\int_{0}^{\pi}\cos\theta\ln\big(R-r\cos\varphi\cos\theta+\sqrt{R^2+r^2-2rR\cos\varphi\cos\theta}\big)\:d\theta
\end{equation}
\begin{equation}
I_3=\int_{0}^{\pi}\cos\theta\ln\big[r(1-\cos\varphi\cos\theta)\big]d\theta
\end{equation}

Now, $I_1$ can be expanded in term of the Legendre polynomials as follows
\begin{equation}
I_1=\int_{0}^{\pi}R\left[1+\left(\frac{r}{R}\right)^2-2\left(\frac{r}{R}\right)\cos\varphi\cos\theta\right]\cdot\sum_{n=0}^{\infty}\left(\frac{r}{R}\right)^{n}P_n(\cos\varphi\cos\theta)d\theta
\end{equation}
where $P_n$ are Legendre polynomials. We take $P_n$ up to $n=2$, which is
\begin{equation}
\begin{aligned}
I_1&=\int_{0}^{\pi}R\left[1+\left(\frac{r}{R}\right)^2-2\left(\frac{r}{R}\right)\cos\varphi\cos\theta\right]\cdot\left[1+\left(\frac{r}{R}\right)P_{1}(\cos\varphi\cos\theta)\right.\\
&\:\:\:\:\:\left.+\left(\frac{r}{R}\right)^{2}P_{2}(\cos\varphi\cos\theta)+O\bigg(\left(\frac{r}{R}\right)^3\bigg)\right]\\
&=R\pi\left[1+\frac{1}{4}\left(\frac{r}{R}\right)^{2}(1+\sin^2\varphi)+O\bigg(\left(\frac{r}{R}\right)^3\bigg)\right]
\end{aligned}
\end{equation}

For $I_2$, we first expand the root term in the integrand by Legendre polynomials (up to $P_2$). It follows that
\begin{equation}
\begin{aligned}
I_2=\int_{0}^{\pi}\cos\theta\ln\left[1-\left(\frac{r}{R}\right)\cos\varphi\cos\theta+\frac{1}{4}\left(\frac{r}{R}\right)^{2}(-\cos^2\varphi\cos^2\theta+1)+O\bigg(\left(\frac{r}{R}\right)^3\bigg)\right]d\theta
\end{aligned}
\label{con:A7}
\end{equation}
By making use of
\begin{equation}
\ln(1-x)=-\left(x+\frac{x^2}{2}+\frac{x^3}{3}+\frac{x^4}{4}+\cdots\right)
\end{equation}
Expand the logarithmic term in the above integrand to the second order $O(x^2)$, then integrate $I_2$, one obtains
\begin{equation}
I_2=-\frac{\pi}{2}\left(\frac{r}{R}\right)\cos\varphi+O\bigg(\left(\frac{r}{R}\right)^3\bigg)
\end{equation}

Finally, integrating $I_3$ directly, we get
\begin{equation}
I_3=-\frac{\pi}{\cos\varphi}(1-\sin\varphi)
\end{equation}
By substituting $I_1,I_2,I_3$ into Equation (\ref{con:A1}), one obtains
\begin{equation}
\begin{aligned}
V=-2\mathcal{G}\sigma\pi R\left\{1-\left(\frac{r}{R}\right)\sin\varphi+\frac{1}{4}\left(\frac{r}{R}\right)^{2}(3\sin^2\varphi-1)+O\bigg(\left(\frac{r}{R}\right)^3\bigg)\right\}
\end{aligned}
\end{equation}
or written as
\begin{equation}
V=-\frac{2\mathcal{G}m_d}{R}\left\{1-\left(\frac{r}{R}\right)\sin\varphi+\frac{1}{4}\left(\frac{r}{R}\right)^{2}(3\sin^2\varphi-1)+O\bigg(\left(\frac{r}{R}\right)^3\bigg)\right\}
\end{equation}
where $m_d=\sigma \pi R^2$ is the mass of the uniform disk.

2. When $r/R>1$, the potential can be directly expanded in $\rho/r$ by means of Legendre polynomials:
\begin{equation}
\begin{aligned}
V=-2\mathcal{G}\sigma\int_{0}^{R}\int_{0}^{\pi}\frac{\rho}{r}\sum_{n=0}^{\infty}\left(\frac{\rho}{r}\right)^{n}P_{n}(\cos\varphi\cos\theta)d\rho d\theta
\end{aligned}
\end{equation}
taking $n=0,1,2,3$, we get
\begin{equation}
\begin{aligned}
V&=-\frac{\mathcal{G}m_d}{r}\left[1-\frac{1}{4}\left(\frac{R}{r}\right)^2\left(1-\frac{3}{2}\cos^2\varphi\right)+O\bigg(\left(\frac{R}{r}\right)^4\bigg)\right]
\end{aligned}
\end{equation}
\section{The average of Equation (5) \label{sec:A2}}
Average $r|\sin(f+\omega)|$ over the mean anomaly $M$:
\begin{equation}
\begin{aligned}
\overline{r|\sin(f+\omega)|}&=\frac{1}{2\pi}\int_{0}^{2\pi}r|\sin(f+\omega)|dM\\
&=\frac{a}{2\pi\sqrt{1-e^2}}\int_{0}^{2\pi}\left(\frac{r}{a}\right)^3|\sin(f+\omega)|df\\
&=\frac{a}{2\pi\sqrt{1-e^2}}\left\{\int_{-\omega}^{\pi-\omega}\frac{(1-e^2)^3}{(1+e\cos f)^{3}}\sin(f+\omega)df-\int_{\pi-\omega}^{2\pi-\omega}\frac{(1-e^2)^3}{(1+e\cos f)^{3}}\sin(f+\omega)df\right\}
\end{aligned}
\end{equation}
For $(1+e\cos f)^{-3}$, we expand it to $e^2$, and then integrate the above equation. This results
\begin{equation}
\begin{aligned}
\overline{r|\sin(f+\omega)|}=\frac{2a(1-e^2)^{5/2}}{\pi}\left[(1+3e^2)-e^2\cos2\omega+O(e^3)\right]
\end{aligned}
\end{equation}
Expanding $(1-e^2 )^{5/2}$ to $e^2$ again yields
\begin{equation}
\overline{r|\sin(f+\omega)|}=\frac{2a}{\pi}\left[\left(1+\frac{1}{2}e^2\right)-e^2\cos2\omega\right]+O(e^3)
\end{equation}

\section{Kuzmin disk} \label{sec:A3}

Kuzmin disk is a classical model for the razor-thin disk galaxy, and it is of infinite radial extent but has finite mass as the surface density decreases fast with radius. The potential-density pairs of Kuzmin disk is given by \citep{2008gady.book.....B}
\begin{subequations}
\begin{align}
V_K(R_k,z) &=\frac{-\mathcal{G}M_K}{\sqrt{R_k^2+(C+|z|)^2}} \label{con:sub1} \\
\Sigma_K(R_k)&=\frac{CM_K}{2\pi(R_k^2+C^2)^{3/2}}\label{con:sub2}
\end{align}
\end{subequations}
where $M_k$ is the total mass of Kuzmin disk, $R_k^2=x^2+y^2$. $C(>0)$ is the radial scale length of the disk. Equation (\ref{con:sub1}) can also be written as
\begin{equation}
V_K=\frac{-\mathcal{G}M_K}{\sqrt{r^2+C^2+2rC | \sin \varphi|}}
\label{con:eq29}
\end{equation}
where $r^2=R_k^2+z^2=x^2+y^2+z^2$ and $\sin\varphi=|z|/r$.

When $r/C<1$, we can expand Equation (\ref{con:eq29}) in $r/C$ using Legendre polynomials. Here we take Legendre polynomial up to $P_2$, and then we get
\begin{equation}
V_K=-\frac{\mathcal{G}M_K}{C}\left[1-\left(\frac{r}{C}\right)\sin\varphi+\left(\frac{r}{C}\right)^2\left(\frac{3}{2}\sin^2\varphi-\frac{1}{2}\right)\right]
\end{equation}
The average of $V_K$ over the mean anomaly $M$ is
\begin{equation}
\begin{aligned}
\overline{V}_K=\frac{\mathcal{G}M_K}{C}&\left\{\left(\frac{a}{C}\right)\frac{2\sin i}{\pi}\left[\left(1+\frac{1}{2}e^2\right)-e^2\cos2\omega\right] \right.\\
&+\left.\left(\frac{a}{C}\right)^2\left[\frac{1}{2}\left(1+\frac{3}{2}e^2\right)\left(1-\frac{3}{2}\sin^2i\right)+\frac{15}{8}e^2\sin^2i\cos2\omega\right]\right\}
\end{aligned}
\label{con:eq31}
\end{equation}
The above equation omits the constant term not involving the orbital elements. By comparing Equation (\ref{con:eq31}) and Equation (\ref{con:eq8}), it is easy to find that they are identical in form and only different in the numerical factor.


\bibliography{sample63}{}
\bibliographystyle{aasjournal}



\end{document}